\documentclass[a4paper,twocolumn,fullpaper]{jpsj3} 
\usepackage{ulem}
\title{
Infrared and Terahertz Spectroscopy 
of Strongly Correlated Electron Systems 
under Extreme Conditions
}
\author{
Shin-ichi \textsc{Kimura}$^1$\thanks{E-mail address: kimura@ims.ac.jp} 
and Hidekazu \textsc{Okamura}$^2$\thanks{E-mail address: okamura@kobe-u.ac.jp}
}
\inst{
$^1$UVSOR Facility, Institute for Molecular Science, and School of Physical Sciences, The Graduate University for Advanced Studies (SOKENDAI), Okazaki, Aichi 444-8585, Japan \\
$^2$Department of Physics, Graduate School of Science, Kobe University, Kobe, Hyogo 657-8501, Japan
}
\recdate{
\today
}
\abst{
Owing to its high brilliance, infrared and terahertz synchrotron radiation (IR/THz-SR) has emerged as a powerful tool for spectroscopy under extreme (i.e., technically more difficult) experimental conditions such as high pressure, high magnetic field, high spatial resolution, and a combination of these.  
The methodologies for pressure- and magnetic-field-dependent spectroscopy and microscopy using IR/THz-SR have advanced rapidly worldwide.  
By applying them in strongly correlated electron systems (SCESs), many experimental studies have been performed on their electronic structures and phonon/molecular vibration modes under extreme conditions.   
Here, we review the recent progress of methodologies of IR/THz-SR spectroscopy and microscopy, and the experimental results on SCESs and other systems obtained under extreme conditions.  
}
\kword{infrared synchrotron radiation, terahertz, strongly correlated electron systems, high pressure, magnetic field}
\begin{document}
\maketitle
%
\section{Introduction}\label{sec:intro}

\begin{figure}[b]
\begin{center}
\includegraphics[width=0.45\textwidth]{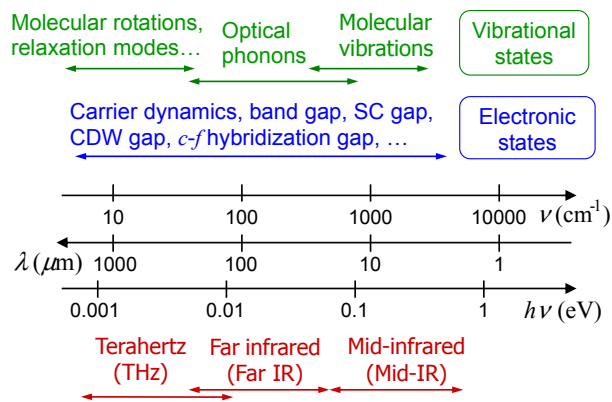}
\end{center}
\caption{
(Color online)
Summary of various phenomena, which have their characteristic frequencies and energies in the IR and THz ranges.
They are indicated relative to the wavenumber ($\nu$) in cm$^{-1}$, wavelength ($\lambda$) in $\mu$m, and photon energy ($h\nu$) in eV.  
Shown at the bottom are the technical terms specifying particular spectral  ranges, which are used in the text.
SC, CDW, and $c$-$f$ represent superconducting, charge density wave, and conduction-$f$ electron, respectively.   
}
\label{fig:IR}
\end{figure}

The physical properties of solids originate from electronic structures near the boundary between occupied and unoccupied states, namely, the chemical potential or Fermi level ($E_{\rm F}$).  To observe such electronic structures, experimental probes with an energy scale corresponding to the temperature at which the physical properties appear must be used.  Light in the infrared (IR) and terahertz (THz, or far-IR) regions has photon energy in the range of 10$^{-4}$ to 1 eV, which roughly corresponds to the temperature range of 1 to $10^4$~K.  Therefore, spectra in the IR and THz regions contain information on the origin of thermodynamic properties.   Figure~\ref{fig:IR} shows a summary of various phenomena whose characteristic energies and frequencies are in the IR and THz regions.

In strongly correlated electron systems (SCESs), novel physical phenomena are often observed at the boundary of the Mott transition, where the on-site Coulomb interaction energy is close to the bandwidth, and also at the quantum critical point, where a magnetic transition occurs at zero temperature.  Examples of such phenomena include a metal-insulator transition, exotic superconductivity, colossal magnetoresistance, and the so-called ``heavy-fermion'' formation.  Investigation of the electronic structure is important in order to understand the origin of these interesting phenomena and to discover even more novel phenomena.  IR/THz spectroscopy is a suitable method of observing changes in the electronic structure responsible for these phenomena induced by changes in temperature, external pressure, and magnetic field.

IR/THz spectroscopy has a long history because IR/THz light can be easily obtained using thermal light sources such as the globar (a SiC rod heated to $\sim$ 1500~K) and the mercury lamp (a discharge lamp with mercury vapor), that utilize the black-body radiation from a heated object.
In the 1960s, Palmer and Tinkham measured the superconducting gap of lead,~\cite{Palmer1968} after which many pioneering works to observe lattice vibrations and magnetic orderings were performed.~\cite{Mitra1968}
In subsequent decades, quasiparticle states of heavy-fermions, high $T_c$-cuprates, and organic superconductors were measured using thermal IR sources.~\cite{Degiorgi1999, Basov2011}
These measurements were generally performed using large samples of several mm sizes.
Minute samples of sub-mm size are more difficult to measure in the IR/THz range with a thermal IR source, since it is not a bright source as will be explained in detail later.
Newly developed materials are usually small at the initial stage.
To study such minute samples using spectroscopy, light must be focused onto a small region.
Therefore, brighter IR/THz light sources are clearly more advantageous for studying them.  This is also the case with IR/THz studies in a restricted sample space, such as high-pressure and high-magnetic-field studies.

\begin{figure}[b]
\begin{center}
\includegraphics[width=0.45\textwidth]{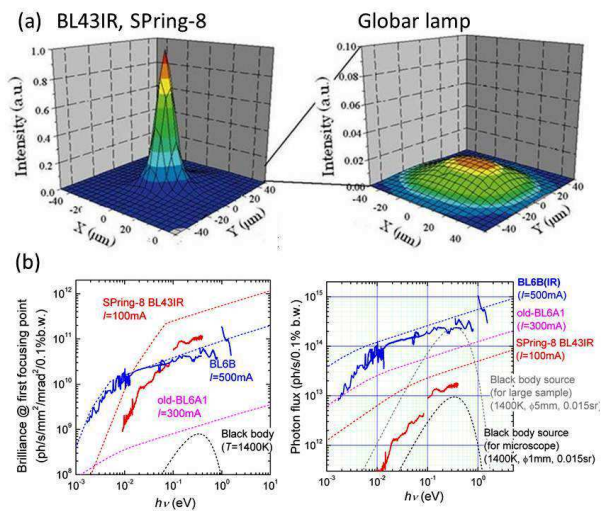}
\end{center}
\caption{
(Color online)
(a) Beam profiles for IR-SR (left) and a globar (right), measured with an IR microscope at the beamline BL43IR of SPring-8.\cite{matunami-2004} [See also Fig.~\ref{fig:inst}(a)].  Here, a 2-$\mu$m-diameter pinhole was scanned along the focal plane of the microscope while recording the integrated intensity over 500--8000~cm$^{-1}$.  The apertures A$_1$ and A$_2$ shown in Fig.~\ref{fig:inst}(a) were not used.  The objective of the IR microscope had a magnification of 8$\times$ and a numerical aperture of 0.5.  For the globar data, the built-in globar of the spectrometer was used, as in Fig.~\ref{fig:inst}(a).
(b) Brilliance (left) and photon flux (right) spectra of the beamline BL6B at UVSOR-II, BL43IR at SPring-8, and a black body light source.  The dashed and solid lines indicate the calculated and measured spectra, respectively.  The measured photon flux was calibrated by a black body source.  The measured brilliance was evaluated using the measured photon flux spectra, the obtained beam size at the first focal position, and the expected divergence.}
\label{IRSR}
\end{figure}
In the 1980s, many synchrotron radiation (SR) facilities were constructed worldwide, and actively used as vacuum-ultraviolet (VUV) and X-ray sources.~\cite{Willmott2011}
SR is characterized by its ability to produce highly parallel and narrow beams of light.
This property can be expressed as having a ``high brilliance''.
Brilliance refers to the number of photons radiated from a unit source area into a unit solid angle, in a unit time for a given spectral bandwidth.~\cite{Willmott2011}
(The term ``emittance'', which is the product of source size and the light beam divergence along one dimension, is also often used.
A high brilliance is usually accompanied by a low emittance.)
SR has high brilliance not only in the VUV/X-ray region but also in the IR/THz region.~\cite{williams-2002}
Examples of beam profiles of the IR-SR from SPring-8 synchrotron and a globar are shown in Fig.~\ref{IRSR}(a).
(For technical details of the measurement conditions, see the caption of Fig.~\ref{IRSR}.)
It is clearly seen that IR-SR can be focused onto a much narrower area with two orders of magnitude higher peak intensity than the globar, owing to the much higher brilliance of the former.

In order to make use of the high brilliance of IR/THz-SR, the first dedicated THz beamline was constructed at the UVSOR Facility of the Institute for Molecular Science in 1984.~\cite{Nanba1986}
Following the successful installation of this beamline, more than 20 IR/THz beamlines are currently in operation or under construction over the world.~\cite{sirup}
The main purpose of these IR/THz beamlines is IR microspectroscopy and imaging with diffraction-limited spatial resolution making use of the high brilliance of SR.\cite{carr-2001}
There are now four IR/THz beamlines in Japan: BL6B for IR/THz microspectroscopy~\cite{Kimura2006} and BL1B for THz coherent synchrotron radiation (CSR) at UVSOR,~\cite{Kimura2012} BL43IR for IR microspectroscopy~\cite{Kimura2001,Kimura2003,Ikemoto2004} and scanning near-field microscopy~\cite{Ikemoto2012} at SPring-8, and BL-15 for IR microspectroscopy at the SR Center of Ritsumeikan University.~\cite{Yaji2008}
The experimental and theoretical spectral distributions of brilliance and photon flux of BL6B at UVSOR and BL43IR at SPring-8 in comparison with a black-body source are shown in Fig.~\ref{IRSR}(b).
The figure indicates that the IR-SRs of both BL43IR and BL6B have much higher brilliance and photon flux than the black-body source.
BL43IR has higher brilliance in the mid-IR region than BL6B, while BL6B has a higher brilliance and a higher photon flux in the THz region.
Many experiments are being performed with both beamlines making use of these properties.

\begin{figure}[b]
\begin{center}
\includegraphics[width=0.45\textwidth]{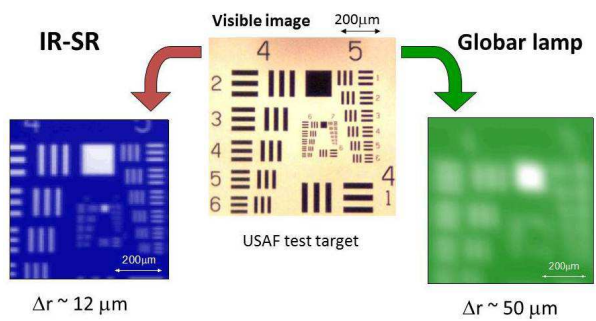}
\end{center}
\caption{
(Color online)
IR mapping image of USAF test target observed using IR-SR and a globar lamp, measured at the beamline BL6B of UVSOR.
The integrated intensity at over 1000-9000~cm$^{-1}$ was recorded while scanning the sample in the focal plane in 5~$\mu$m steps.
The objective of the IR microscope had a magnification of 8$\times$ and a numerical aperture of 0.5.
The apertures A$_1$ and A$_2$ shown in Fig.~\ref{fig:inst}(a) were not used.
The spatial resolution ($\Delta r$) was evaluated as 12~$\mu$m using IR-SR and 50~$\mu$m using the globar lamp.
}
\label{USAF}
\end{figure}
To further demonstrate the performance of IR-SR, IR images of a U.S. Air Force (USAF) test target measured with IR-SR and a globar are shown in Fig.~\ref{USAF}.~\cite{USAF}  The test target, on which a metal pattern is formed on a quartz substrate, was used for the evaluation of spatial resolution.  The test target was measured on the focal plane of an IR microscope at the beamline BL6B of UVSOR.  (For technical details of the measurement conditions, see the caption of Fig.~\ref{USAF}.)  The spatial resolution was evaluated as 12~$\mu$m using IR-SR and 50~$\mu$m using a globar lamp.  The spatial resolution using IR-SR is close to the diffraction limit of about 10~$\mu$m at 1000~cm$^{-1}$, demonstrating the high potential of IR-SR for IR microspectroscopy.

As discussed above, the most important advantage of IR-SR over the thermal IR sources is its high brilliance.
Accordingly, the most common and popular application of IR-SR is to microspectroscopy with a high spatial resolution, in either a spatially resolved study within a large sample or a study of a small sample.\cite{wirms-2011}
IR microspectroscopy is currently one of the most common analysis methods for various materials in many fields of science and industry.\cite{griffiths-2009}
A wide range of commercial IR microspectroscopy instruments, designed for mid-IR spectral range with a thermal IR source, are available.\cite{griffiths-2007}
However, with the use of IR-SR, one can further improve the spatial resolution of IR microspectroscopy.\cite{Nasse2011}
In addition, in the THz and far-IR ranges, the brilliance of the thermal source is even lower, and microspectroscopy with a high spatial resolution is difficult without IR-SR.
Even without a microscope, the high brilliance and low emittance of IR-SR are useful, for example, for magneto-optical study with a superconducting magnet, as discussed in \S\ref{sec:mag} in detail, and for the spectroscopy of molecular vibrations and rotations with a very high spectral resolution. (A spectral resolution as high as 0.0008~cm$^{-1}$ has been obtained with IR-SR for gaseous samples.~\cite{Albert2011})

In addition to the high brilliance and low emittance, there are other useful  features of IR-SR.
First, SR is a very broadband IR source, covering the entire THz-IR ranges shown in Fig.~\ref{fig:IR}.
In addition, SR provides not only linear polarization in the orbital plane of the electron beam but also circular polarization (elliptical polarization, to be precise) out of the orbital plane.
Therefore, linear/circular dichroism experiments can be performed over a very wide spectral range, as discussed in detail in \S\ref{sec:mag}.
(To study dichroism with a thermal IR source, in contrast, additional optical elements such as a quarter wavelength plate are necessary, but they often limit the available spectral width.)

Another important IR technique using polarization is ellipsometry.\cite{Schubert2004}
Ellipsometry is a powerful method that can determine the dielectric function of various metals and dielectrics without Kramers-Kronig analysis (discussed in \S\ref{sec:inst}).
It is also an important tool for the characterization of thin films and multilayered materials.
Nowadays, ellipsometers are commercially available~\cite{JAW} and are widely used with thermal sources, for example, in the industry for material characterization and on-line quality control.
However, with a thermal IR source, ellipsometry is generally more difficult in the far-IR and THz ranges with long wavelengths, since it is more difficult to have a well-collimated beam. 
Therefore, an ellipsometry apparatus for use in the far-IR region has been developed using IR-SR at a few SR facilities, including ANKA (Karlsruhe, Germany),~\cite{Bernhard2004} BESSY II (Berlin, Germany),~\cite{Hofmann2006} and NSLS (Brookhaven, USA).
The ellipsometry using IR/THz-SR allows us to perform precise far-IR ellipsometry measurement in the photon energy range down to 8~meV on thin films and multilayers and even on small single crystals of various materials in a wide temperature range from 5 to 500~K.~\cite{Schubert2004}
Bernhard {\it et al.} measured the $c$-axis $\sigma(\omega)$ spectra of a high-$T_c$ cuprate~\cite{Bernhard1998} and of an iron-based superconductor~\cite{Dubroka2008}, the dynamical response at the LaAlO$_3$/SrTiO$_3$ interface~\cite{Dubroka2010}, and others.

Furthermore, SR is a pulsed source since it is radiated from bunches of electrons, rather than from a continuous beam of electrons.  
The temporal width of the SR pulse is determined by the length of the electron bunch.   It varies among different SR facilities, but is generally between 20 and 500~psec.  Therefore, transient phenomena with similar time scales have been studied by time-resolved IR-SR spectroscopy.  They are mainly interband phenomena, such as the recombination of electron-hole pairs and Cooper pairs after photo-excitation by a pump laser pulse.  Although pulsed IR lasers with much shorter pulse widths are currently available, they are basically monochromatic or narrow-band sources.  In contrast, as mentioned above, SR is an extremely broad-band source.  In addition, CSR, whose intensity is several orders of magnitude higher than the normal SR in the THz range (discussed in \S\ref{sec:CSR}), also has similar features.  Examples of materials that have been studied by time-resolved, broadband IR-SR spectroscopy include superconducting metals such as Pb,~\cite{carr-2000,carr-2005} semiconductors such as GaAs,~\cite{carr-1993,carr-1999b,carr-2002,carr-2003} and organic compounds,~\cite{tahara-2001} which are made in laser-pump, IR-SR-probe experiments.  

In this article, recent IR/THz-SR studies of various materials under extreme conditions are reviewed.
Here, the materials discussed are mainly SCES compounds such as rare-earth (``heavy-fermion'') compounds, transition-metal compounds, and organic compounds.
``Extreme conditions''  simply refer to spectroscopic studies at high pressure and high magnetic fields, microspectroscopy and imaging at a high spatial resolution, and combinations of these conditions.
Although the magnitudes of pressure and magnetic field used in these works may not be extremely large for the high-pressure and high-field communities, it is not easy to perform IR spectroscopy under these conditions without high-brilliance IR-SR.
We intended to cover a wide range of IR-SR studies reported by many workers, particularly for high-pressure studies, to showcase the usefulness of IR-SR.
At the same time, we also intended to discuss selected works of our own in more detail.

The text below is organized as follows.
In \S\ref{sec:inst}, we firstly review the basic instrumentation at an IR-SR beamline and then discuss the data analysis for deriving the optical functions from the measured spectra.
In \S\ref{sec:pressure}, we review high-pressure studies of phonons and electronic structures in SCES, performed with IR-SR and a diamond anvil cell.
In \S\ref{sec:mag}, we review magneto-optical and magnetic dichroism studies performed with IR-SR and a superconducting magnet.
In \S\ref{sec:imaging}, we review IR  microspectroscopy of micro-patterned graphene and carbon nanotube devices with a very high spatial resolution, as a good example where the high brilliance of IR-SR is fully utilized.
Then, we review the IR spectroscopic imaging of organic superconductors and magnetoresistive oxide.
In \S\ref{sec:CSR}, we discuss the future directions of research with IR-SR and other accelerator-based sources, particularly the CSR.
In \S\ref{sec:conc}, we give the conclusions.  

\section{Basic Instrumentation and Data Analysis\cite{okamura-2012c}}\label{sec:inst}
\begin{figure}[b]
\begin{center}
\includegraphics[width=0.45\textwidth]{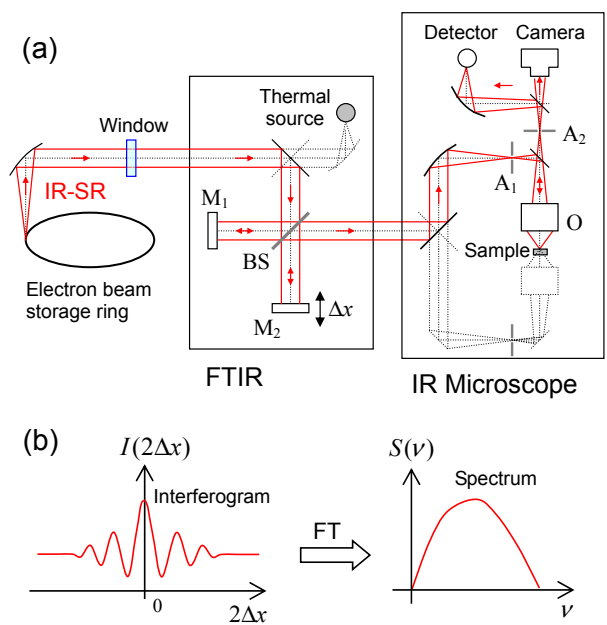}
\end{center}
\caption{
(Color online) 
(a) Schematic diagram of instrumentation used for IR microspectroscopy at an IR-SR beamline.
The symbols indicate the following.  FTIR: Fourier transform IR spectrometer; 
M$_1$: fixed mirror; M$_2$: movable mirror that is scanned to produce the optical path difference 2$\Delta x$; 
BS: beam splitter; 
A$_1$ and A$_2$: apertures (pinholes) in the IR microscope to reduce the effective beam size at the sample position; 
O: reflecting objective (also called Schwarzschild or Cassegrain mirror).
In the FTIR, the optical path for using a built-in thermal source is indicated by dotted lines.
In the microscope, the optical path for the  reflectance study is indicated by the solid lines, and that for the transmittance study by the dotted lines.
Note that, with high-brilliance IR-SR, a nearly diffraction-limited spot size may be obtained without the apertures A$_1$ and A$_2$.
(b) Data processing in FTIR.
The interferogram measured as a function of the optical path difference, 
$I(2\Delta x)$, is Fourier-transformed to obtain the spectrum $S(\nu)$.  
}
\label{fig:inst}
\end{figure}
Figure~\ref{fig:inst}(a) schematically shows an example of the instrumentation used for microspectroscopy at an IR-SR beamline.  The IR-SR from the electron beam is collected by appropriate optics and carried to the experimental station through evacuated beam transport.
In general, the IR-SR beam is first input to a Fourier transform IR spectrometer (FTIR),~\cite{griffiths-2007} which is basically a Michelson interferometer.
The output beam from the FTIR is carried to an IR microscope, focused onto the sample, and then the reflected or transmitted beam is input to a detector.
Since SCES samples generally have a very low transmittance in the IR range, the reflectance (or equivalently the reflectivity) spectrum $R(\omega)$ is commonly studied.
In this case, a gold or silver film is also studied as reference to normalize $R(\omega)$.
[The $R(\omega)$ of gold and silver in the IR is almost unity.]
The signal output from the detector is used to record the interference pattern called the interferogram, as depicted in Fig.~\ref{fig:inst}(b).
The Fourier transform of the interferogram gives the spectrum of the detected IR beam.
FTIR has a multiplex advantage (detecting a wide frequency range at once) and a throughput advantage (high throughput due to the absence of a slit) over the grating-based spectrometer, and hence widely used for infrared studies of materials.  
FTIR usually contains a built-in, thermal IR source (a globar or mercury lamp), as shown in Fig.~\ref{fig:inst}(a), which is used when not using the IR-SR.  

For spatially resolved measurements such as the mapping and high pressure experiments discussed later, an IR microscope~\cite{griffiths-2007,griffiths-2009} is usually used, as schematically shown in Fig.~\ref{fig:inst}(a). 
In this case, the output from the FTIR is input to the microscope and tightly focused onto the sample with a reflecting objective (called the Schwarzschild or Cassegrain  mirror), which has a high magnification (generally 8$\times$ to 32$\times$).
Owing to the high brilliance of IR-SR, discussed in \S\ref{sec:intro}, it is possible to obtain a beam spot size of the order of the wavelength ({\it i.e.}, diffraction-limited spot size) at the sample position, without using the apertures A$_1$ and A$_2$ in Fig.~\ref{fig:inst}.
On the other hand, when the thermal IR source in the FT-IR is used, these apertures must be used to obtain a small beam, since the source size is large ($\sim$ 1~cm).
Even when the source size is reduced to 1~mm by another aperture, the emittance in a typical FTIR is roughly 1 to 2 orders of magnitude larger than that of IR-SR.~\cite{footnote}
Hence, the light intensity that reaches the sample through these apertures should be very small.
This again shows that the IR-SR is an excellent source for IR microscopy compared with the conventional, thermal IR source.

For low-temperature studies, the sample is mounted on a cryostat held in vacuum.
For high-magnetic-field studies, the sample and cryostat are inserted into the bore of a superconducting magnet.
Since the IR radiation is strongly absorbed by H$_2$O and CO$_2$ in air, the optical paths for the IR beam are either vacuum-pumped, purged with N$_2$ gas, or dry air.

As mentioned above, a commonly measured quantity in the study of SCES samples is $R(\omega)$ under a near-normal-incidence condition.
In this situation, the complex reflectivity of the electric field component of the electromagnetic radiation may be expressed as,~\cite{DresselGruner,Wooten}
\begin{equation}
\hat{r}(\omega)= \frac{1 - \hat{n}(\omega)}{1 + \hat{n}(\omega)}
= r(\omega) e^{i \theta(\omega)}. 
\label{eq:fresnel1} 
\end{equation}
Here, $r(\omega)=\sqrt{R(\omega)}$ and $\theta(\omega)$ are the amplitude reflectivity and phase shift of the electric field, respectively, and $\hat{n}(\omega)$ is the complex refractive index of the sample.
$R(\omega)$ is also commonly expressed as 
\begin{equation}
R(\omega)= \left|
\frac{1 - \hat{n}(\omega)}{1 + \hat{n}(\omega)}
\right|^2 =
\frac{(1-n)^2+k^2}{(1+n)^2+k^2},
\label{eq:fresnel2}
\end{equation}
where $\hat{n}(\omega)= n(\omega) + ik(\omega)$.
If both $n(\omega)$ and $k(\omega)$ are known, or if both $r(\omega)$ and $\theta(\omega)$ are known, one can derive all the other optical functions such as the dielectric function and optical conductivity.
However, only $r(\omega)=\sqrt{R(\omega)}$ is obtained in the actual reflectance measurement.
Therefore, to obtain the optical functions from measured $R(\omega)$, Kramers-Kronig (KK) analysis and Drude-Lorentz (DL) spectral fitting have been widely used.

In the KK analysis of reflectance data, the KK relation between $r(\omega)$ and $\theta(\omega)$, namely,
\begin{equation}
\theta(\omega) = - \frac{2\omega}{\pi} P \int_0^\infty 
 \frac{{\rm ln} r(\omega^\prime)}{\omega^{\prime 2} - \omega^2}
d\omega^\prime,
\label{eq:KK}
\end{equation}
where $P$ denotes the principal value and $r(\omega)=\sqrt{R(\omega)}$, as already mentioned earlier, is used.
Namely, if $r(\omega)$ is known at all values of $\omega$, $\theta(\omega)$ may be obtained by carrying out the integration in eq.~\ref{eq:KK}.
In practice, $R(\omega)$ can be measured only over a finite frequency range.
Accordingly, appropriate extrapolations are used at both low- and high-frequency ends of the measured $R(\omega)$.

\begin{figure}[b]
\begin{center}
\includegraphics[width=0.4\textwidth]{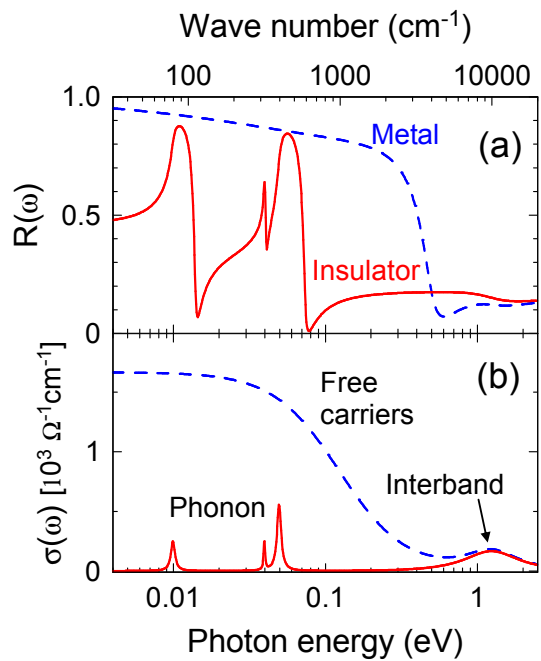}
\end{center}
\caption{
(Color online)
(a) $R(\omega)$ spectra calculated from the model dielectric function of eq.~(\ref{eq:DL}) with two sets of different parameters.
The dashed curve (``Metal'') simulates a metal with one Drude oscillator ($\omega_p$=10000~cm$^{-1}$, $\gamma$=1000~cm$^{-1}$), one Lorentz oscillator ($\omega_0$=$\gamma$=10000~cm$^{-1}$), and $\varepsilon_\infty$ =5.
The solid curve (``Insulator'') simulates an insulator with three Lorentz oscillators for phonons ($\omega_0$=80, 320, and 400~cm$^{-1}$ with $\gamma$=6, 8, and 30~cm$^{-1}$, respectively.) and also the same  Lorentz oscillator at 10000~cm$^{-1}$, as in the metal case.
(b) $\sigma(\omega)$ spectra given by the same parameters as in (a), showing the  characteristic spectral components due to free carriers (Drude response), phonons, and interband transition.
}
\label{fig:DL}
\end{figure}

In contrast, in the DL fitting analysis of $R(\omega)$ data, one uses a model dielectric function expressed as a sum of Drude and Lorentz oscillators, which represent free and bound electrons, respectively:~\cite{DresselGruner,Wooten}
\begin{equation}
\hat{\varepsilon}(\omega)=\varepsilon_\infty + 
\sum_{j} \frac{\omega_{p,j}^2}
{\omega_{0,j}^2 - \omega^2 - i \omega \gamma_j} 
\label{eq:DL}
\end{equation}
where $\omega_p$, $\omega_0$, and $\gamma$ are the plasma frequency, natural frequency, and scattering rate, respectively.
$j$ denotes the $j$th oscillator, and $\omega_0$=0 for a Drude oscillator.
$\varepsilon_\infty$ is a parameter representing the polarizability of higher-lying interband transitions.
Using the relation $\hat{n}(\omega)=\sqrt{\hat{\varepsilon}(\omega)}$, $\hat{\varepsilon}(\omega)$ given by a set of input parameters is substituted into eq.~(\ref{eq:fresnel1}) to calculate $R(\omega)$.
 Figure~\ref{fig:DL}(a) shows two $R(\omega)$ spectra calculated with two sets of parameters, simulating a metal (dashed curve) and an insulator (solid curve).
The metal case has a high reflectivity (plasma reflection) due to a Drude oscillator (free carriers) and an interband electronic transition centered at 4000~cm$^{-1}$.
The insulator case has three marked structures in the far IR range due to optical phonons.  (The actual parameters used are indicated in the caption.)
Such calculations are repeated while adjusting the parameters to minimize the difference between the calculated and measured $R(\omega)$ spectra.
This is  usually performed by the least-squares fitting method.
Using the best-fit parameters, all the optical functions may be calculated.
The $\sigma(\omega)$ spectra corresponding to the $R(\omega)$ spectra in Fig.~\ref{fig:DL}(a) are shown in Fig.~\ref{fig:DL}(b).
In the metal case, $\sigma(\omega)$ rises toward zero frequency, which is a characteristic spectral response of free carriers.
In the insulator case, sharp peaks due to phonons are observed, generally below 1000~cm$^{-1}$.
The interband transition of electrons also causes a peak in $\sigma(\omega)$.

\section{IR Spectroscopy under High Pressure}\label{sec:pressure}

\begin{figure}[b]
\begin{center}
\includegraphics[width=0.45\textwidth]{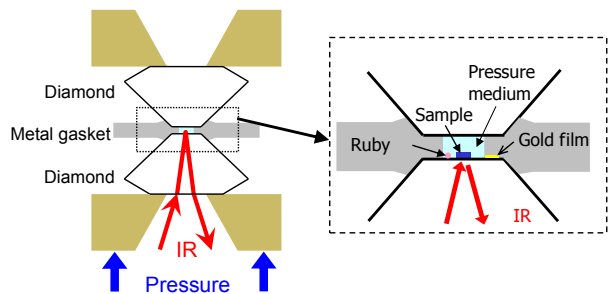}
\end{center}
\caption{(Color online) 
Schematic sketch of a sample setting in a DAC for an IR reflectance study.
Ruby is used as a pressure sensor through its fluorescence, and a gold film is used as the reference of reflectance.  
The surface of the gasket or the culet surface of an ``empty'' DAC, as shown in the inset of Fig.~\ref{fig:phonon}(a), has also been used as the reference of reflectance.     
}
\label{fig:DAC}
\end{figure}
The application of external pressure has been a useful and popular experimental technique in condensed matter physics.
Since the applied pressure on a material reduces the interatomic distance in the material, it can be used to tune material parameters such as the electron hybridization and bandwidth, the ionic radii of the atoms, and the electron and phonon densities of states.
Accordingly, an external pressure often induces remarkable changes in the crystal structure and/or the electronic structures of materials.  
For example, structural phase transitions, metal-insulator transition, superconductivity, and quantum critical phenomena have been observed under high pressure.  
To study the pressure-dependent electronic structures, IR spectroscopy is particularly useful, since other common spectroscopic methods such as photoemission and tunneling techniques are technically difficult to perform under pressure.   
To perform an optical study of a material under high pressure, a diamond anvil cell (DAC) has been commonly used.~\cite{HP1,HP2}  
Figure~\ref{fig:DAC} shows an example of sample setting in a DAC for an IR reflectance study.\cite{okamura-2010}    
A sample is confined by the culet faces of a pair of diamond anvils, a metal gasket with a through hole, and a pressure-transmitting medium. 
Since diamond is mostly transparent in the IR and THz ranges, optical transmission or reflection experiments can be performed.  
The available sample space is, however, very limited since the culet diameter of the anvil should generally be smaller than 1~mm to produce a pressure of 5--7~GPa or above.  
As a result, only small samples, generally smaller than 0.5~mm, can be used in a DAC.

To perform IR spectroscopy in the restricted sample space in a DAC, the high brilliance of IR-SR, discussed in \S\ref{sec:intro}, is apparently advantageous.    
Therefore, since the initial stage of IR-SR research in the 1980s, the high-pressure study using DAC has become one of the major applications of IR-SR.   
Early works included the far-IR phonon studies of alkali halides across the pressure-induced structural phase transition (B1-B2 transition) at UVSOR,~\cite{nanba-1989} and the vibrational spectroscopy of dense molecular solids such as solid hydrogen and ice at NSLS.~\cite{hemley-1997}

Since the above works, high-pressure IR studies have been performed at other SR facilities as well.
The major topic among them is probably the metal-insulator transition in various materials such as transition-metal oxides and chalcogenides ($d$ electron systems), organic compounds ($p$ electron systems), and rare-earth compounds ($f$ electron systems). 
Other materials physics topics studied by IR-SR spectroscopy under high pressure include the charge-density-wave (CDW) states, the localized-delocalized transitions and crossovers of $f$ electrons, the phonon states in semiconductor microcrystals and multiferroic compounds, and the electronic structure of elemental metals such as Fe and Yb.
These high-pressure studies are reviewed below, after a brief introduction of the instrumentation and data analysis.
In \S\ref{sec:no-SR}, in addition, some high-pressure results obtained with the thermal IR sources are briefly discussed, but the results discussed in the other sections were obtained with IR-SR.
Other major high-pressure applications of IR-SR, e.g., to the vibrational spectroscopy of dense molecular solids~\cite{hemley-1997} and minerals, are not discussed here.

\subsection{Instrumentation and data analysis for high pressure IR-SR studies}\label{sec:DL}

Several SR facilities have developed high-pressure IR spectroscopy apparatuses in their IR beamlines.
At these beamlines, an IR microscope is commonly used, as already discussed in \S\ref{sec:intro}, to accurately measure the $R(\omega)$ of a small sample loaded into a DAC.
For low-temperature studies, the DAC is mounted on a cryostat.
Regarding the diamond anvil, a Type IIa diamond is suited for IR studies, since it is transparent over a wide spectral range from the far IR to UV, except for a strong two-phonon absorption band at approximately 0.235--0.285~eV (1900--2300~cm$^{-1}$) range.\cite{okamura-2010} 
Type Ia diamonds are less expensive, but usually have an additional strong absorption below 0.15~eV (1200~cm$^{-1}$) due to nitrogen impurities.
If the low energy range below 0.2~eV is of interest, the use of a Type IIa diamond is highly desired.
The pressure in a DAC is usually monitored by the ruby fluorescence method.~\cite{mao-1978,Rogan1992}  
Namely, a small ruby piece is loaded into the DAC with the sample, and its fluorescence (R1 line) is measured with another source such as a green (532~nm) laser.
The pressure is then estimated from the known wavelength-pressure relation of the R1 line.

Various materials have been used as the pressure-transmitting medium for DAC.~\cite{tateiwa-2007,klotz-2009,klotz-2012} 
They include inert gases such as He and Ar, liquids such as mixed methanol-ethanol, glycerin, and Daphne oil,~\cite{murata-2008} and powders of soft solids such as KBr and KCl.
(Powders form a pellet when pressed in a DAC.)
The inert gases can produce the highest hydrostatic pressure, but they require cryo-loading.
The solid media produce relatively lower hydrostatic pressure, but their handling is much easier.
They also allow a close and clean contact between the diamond and the sample, which is important for reflectance studies.
A liquid medium has also been used for reflectance studies to obtain a higher hydrostatic pressure, but its use requires more caution than the solid medium case.  

When $R(\omega)$ of a sample is studied in a DAC, it is measured at the sample/diamond interface. 
Care must be taken, therefore, in its interpretation and analysis owing to the large refractive index of diamond, which is about 2.4 in the IR range.
In Fresnel's formula in eqs.~(\ref{eq:fresnel1}) and (\ref{eq:fresnel2}), the factor 1 (refractive index of vacuum) must be replaced by 2.4.
It is therefore clear that $R(\omega)$ measured in a DAC may markedly differ from that in vacuum.  
In addition, the presence of diamond brings an extra phase shift, $\Delta \theta(\omega)$, into the KK relation of eq.~(\ref{eq:KK}).\cite{plaskett-1963}   
Although the functional form of $\Delta\theta(\omega)$ is usually unknown,~\cite{plaskett-1963} methods of the KK analysis of $R(\omega)$ measured in a DAC have been proposed and successfully used.\cite{klehe-2000,pashkin-2006,okamura-2012b}   
One may also use DL spectral fitting, discussed in \S\ref{sec:inst}, to obtain optical functions, where the diamond refractive index is taken into account using eqs.~(\ref{eq:fresnel1}) and (\ref{eq:fresnel2}).

\subsection{Phonon spectroscopy under high pressure} 

For insulators, semiconductors, and metals with sufficiently low density of free carriers, optical phonons may cause distinct structures and peaks in the optical spectra in the far-IR range.
By studying the frequencies and spectral shape of phonon peaks, information on the vibrational, dielectric, and thermoelastic properties can be obtained.
In addition, the phonon spectrum is sensitive to changes in the symmetry of the crystal structure.
IR spectroscopy is complimentary to Raman spectroscopy, but they have different selection rules for phonons.
By studying the phonon spectrum under pressure, for example, information about pressure-induced structural phase transition, and the associated changes in crystal symmetry, may be obtained.
Some works along this line are reviewed below.

The far-IR phonon absorption spectra of bulk KI~\cite{nanba-1989} and other alkali halides~\cite{nanba-1990}, microcrystals of NaCl\cite{nanba-1994}, and CdS~\cite{nanba-1997} under high pressure were studied by Nanba {\it et al.}
For KI and other bulk alkali harides, they observed changes in the phonon spectrum upon the B1-B2 structural phase transition from the rocksalt to the CsI structure.
They also obtained the mode Gruneisen parameter of a TO phonon in two phases.
For NaCl microcrystals, they observed different phonon frequencies and different shifts with pressure for surface phonons from spherical and cubic microcrystals, and also from bulk phonons.\cite{nanba-1994}   
For CdS microcrystals, they studied the evolution of phonon frequencies and spectral shape with pressure, and found that smaller microcrystals were ``stiffer'' in terms of phonon properties.\cite{nanba-1997}  

\begin{figure}[b]
\begin{center}
\includegraphics[width=0.46\textwidth]{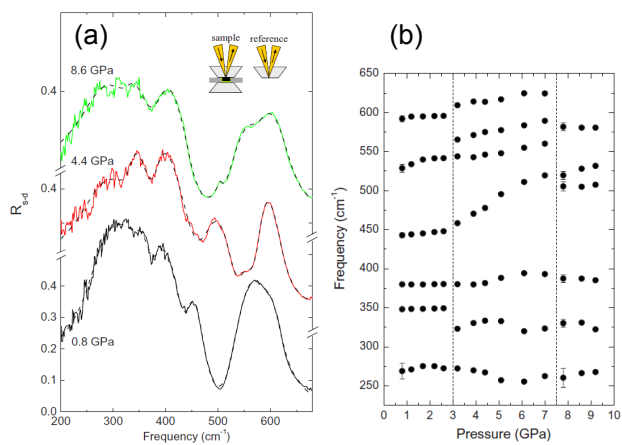}
\end{center}
\caption{
(Color online) 
Far-IR phonon spectroscopy of multiferroic BiFeO$_3$ under high pressure at room temperature reported by Haumont {\it et al.}~\cite{kuntscher-2009}  (a) Reflectance spectra at three pressures, which correspond to different crystal structures.  The broken curves show the results of spectral fitting analysis.   (b) Pressure evolution of the phonon frequencies given by the fitting.  The vertical lines at 3 and 7.5~GPa indicate the pressures where structural phase transitions occur.  
Figure reprinted with permission from Haumont {\it et al.}, Phys. Rev. B {\bf 79}, 184110, 2009.  Copyright 2009 by the American Physical Society.  
}
\label{fig:phonon}
\end{figure}
The pressure evolution of phonon spectra has also been studied with IR-SR for other bulk materials such as ZnO~\cite{dumas-2011}, HfW$_2$O$_8$~\cite{chen-2001}, BiFeO$_3$~\cite{kuntscher-2009}, and $R$MnO$_3$ ($R$=Y, Ho, Lu)~\cite{gao-2011}.
ZnO is a well-known wide-band-gap semiconductor having potential for practical applications, and its thermal properties are also of interest.~\cite{dumas-2011}
HfW$_2$O$_8$ exhibits a negative thermal expansion, and its phonon properties with lattice compression under pressure is of interest.~\cite{chen-2001}
Multiferroic BiFeO$_3$ and $R$MnO$_3$ have recently attracted much interest, since the coupling between magnetic and dielectric properties result in novel physical properties.\cite{kuntscher-2009,gao-2011}  
These compounds show structural phase transitions under high pressure, and changes in the phonon spectrum and the pressure shift of the phonon frequencies were studied in detail.
An example of far-IR reflectance data for BiFeO$_3$,~\cite{kuntscher-2009} which shows both magnetic order and ferroelectricity in the same phase, is shown in Fig.~\ref{fig:phonon}.
The three spectra correspond to three phases with different crystal structures (see the figure caption).
It is seen that some phonon peaks show different responses in the three phases with different crystal structures.

\subsection{Pressure-induced metal-insulator transitions in oxides, chalcogenides, spinels, and others} 

Metal-insulator transition (MIT) is one of the most studied topics in modern condensed matter physics.~\cite{imada-1998}
A large change in the electrical resistivity observed upon MIT is usually accompanied by the appearance or disappearance of energy gap in the density of states.
The application of external pressure is particularly useful for studying MIT in SCES.
In SCES, as mentioned in \S\ref{sec:intro}, the on-site electron correlation ($U$) is on the same order as the electron bandwidth ($W$), and the physical properties of such materials are strongly affected by the balance between them.
An external pressure generally increases $W$, and therefore changes such balance, which may result in an MIT and other interesting phenomena.
A Mott insulator is an example of such a system, where an insulating ground state and the localization of electrons results from a large $U/W$, and the tuning of $U/W$ with external pressure often causes a transition to metallic state and other marked changes in their properties.
Optical spectroscopy has been very useful in studying the evolution of an energy gap in SCES,~\cite{Basov2011} since it can probe the energy-dependent response of electrons.
Below, high-pressure IR-SR studies of MIT on V oxides, 
Ni pyrites, transition-metal spinels, low-dimensional conductors, and other systems are reviewed.    

\begin{figure}[b]
\begin{center}
\includegraphics[width=0.45\textwidth]{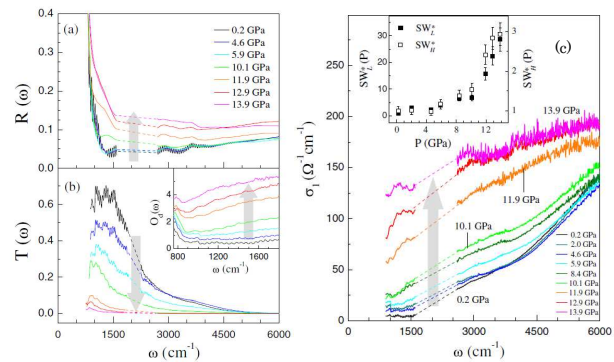}
\end{center}
\caption{
(Color online) 
Optical spectra of VO$_2$ at room temperature under high pressure reported by Arcangeletti {\it et al.},~\cite{arcangeletti-2007} measured on pressed powder samples in a DAC.   (a), (b), and (c) show the reflectance, transmittance, and optical conductivity, respectively.  
Figure reprinted with permission from Arcangeletti {\it et al.}, Phys. Rev. Lett {\bf 98}, 196406, 2007.  Copyright 2007 by the American Physical Society.  
}
\label{fig:VO2}
\end{figure}
Vanadium oxides with various compositions such as VO$_2$, V$_2$O$_3$, and V$_3$O$_5$ are well-known Mott-Hubbard insulators,~\cite{imada-1998,Basov2011} with an energy gap between the two 3$d$ bands split due to the electron correlation.
At ambient pressure, they become metallic above a transition temperature $T_c$.
In the insulating state, they also show pressure-induced transition to a metallic state.
The evolutions of the energy gap and carrier dynamics upon these pressure-induced MIT have been studied with DAC and IR-SR. 
Figure~\ref{fig:VO2} shows an example of a high-pressure IR-SR study of VO$_2$ reported by Arcangeletti {\it et al.}~\cite{arcangeletti-2007}
VO$_2$ at ambient pressure has $T_c=$~340~K, and is in the insulating state at room temperature.
The pressure evolutions of both reflectance and transmittance spectra were measured on a pressed powder sample in DAC, as shown in Figs.~\ref{fig:VO2}(a) and \ref{fig:VO2}(b), respectively, and the $\sigma(\omega)$ spectra in Fig.~\ref{fig:VO2}(c) were obtained from them.  
At a low pressure of 0.2~GPa, $\sigma(\omega)$ has a clear energy gap with an onset of $\sigma(\omega)$ near 1600~cm$^{-1}$ ($\hbar\omega=$~0.2~eV).
With increasing pressure, it is clearly seen that the onset shifts to a lower energy, and the rising portion of $\sigma(\omega)$ above the onset gradually increases up to 10.1~GPa.
From 10.1 to 11.9~GPa, there is a much larger increase, which demonstrates a transition to metallic state at approximately 10~GPa.
A similar study was also performed on the Cr-doped system V$_{1-x}$Cr$_x$O$_2$,~\cite{marini-2008} and on other vanadium oxides with different compositions, i.e., V$_2$O$_3$ with $T_c=$~220~K~\cite{lupi-2012} and V$_3$O$_5$ with $T_c=$~420~K.~\cite{baldassare-2007}
Similar works have been reported on La$_{1-x}$Ca$_x$MnO$_{3-y}$,\cite{sacchetti-2006} and BaVS$_3$,\cite{mihaly-2005} where the pressure evolution of energy gap have been studied with IR-SR.

\begin{figure}[b]
\begin{center}
\includegraphics[width=0.35\textwidth]{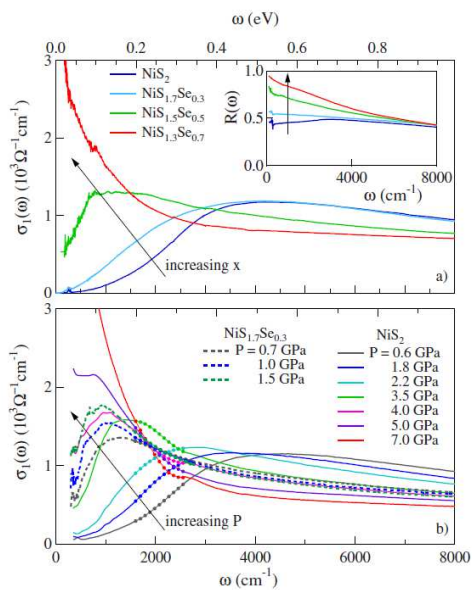}
\end{center}
\caption{
(Color online) 
Optical conductivity [$\sigma_1(\omega)$] spectra of a) NiS$_{2-x}$Se$_x$ at ambient pressure, and b) those of NiS$_2$ and NiS$_{1.7}$Se$_{0.3}$ under high pressure, reported by Kunes {\it et al.}~\cite{kunes-2010}
With increasing Se content and with increasing external pressure, $\sigma_1(\omega)$ shows a similar evolution and the suppression of the energy gap.  
Figure reprinted with permission from Kunes {\it et al.}, Phys. Rev. B {\bf 81}, 035122, 2010.
Copyright 2010 by the American Physical Society.  
}
\label{fig:NiS2}
\end{figure}
NiS$_2$ pyrite at ambient pressure is a charge transfer insulator, where the energy gap is located between the upper 3$d$ band and the S 3$p$ band.
The onset of charge transfer excitation gives the magnitude of energy gap.
A transition to a metallic state may be induced by external pressure or by the chemical substitution of Se for S (chemical pressure).  
In both cases, the energy gap is suppressed owing to an increase in $W$ caused by the crystal lattice contraction, and has been studied by IR-SR spectroscopy.~\cite{perucchi-2009,kunes-2010,perucchi-2011}  
Figure~\ref{fig:NiS2} shows data reported by Kunes {\it et al.}~\cite{kunes-2010}, which compares the evolution of $\sigma(\omega)$ between the Se doping case and the external pressure case.
It is seen that the energy gap is progressively suppressed with increasing Se fraction and pressure.
The evolutions of $\sigma(\omega)$ are qualitatively similar.  

Compounds having the spinel structure with the chemical formula $AB_2C_4$ have attracted many interest recently, partly owing to a magnetic frustration that may act among $B$ ions if the magnetic interaction is antiferromagnetic.
Recently, two spinel compounds, namely, CuIr$_2$S$_4$~\cite{chen-2005} and LiV$_2$O$_4$~\cite{irizawa-2011} have been studied by IR-SR spectroscopy under high pressure.
Both these compounds are metallic at ambient pressure, but with increasing pressure, they progressively become insulating (semiconducting).
This behavior is rather unusual, since many other compounds, including the Mott insulators discussed above, become more metallic with increasing pressure.
The pressure evolution of the unique electronic structures of these compounds were studied in detail from the $\sigma(\omega)$ spectra measured under pressure.~\cite{chen-2005,irizawa-2011}   

\begin{figure}[b]
\begin{center}
\includegraphics[width=0.32\textwidth]{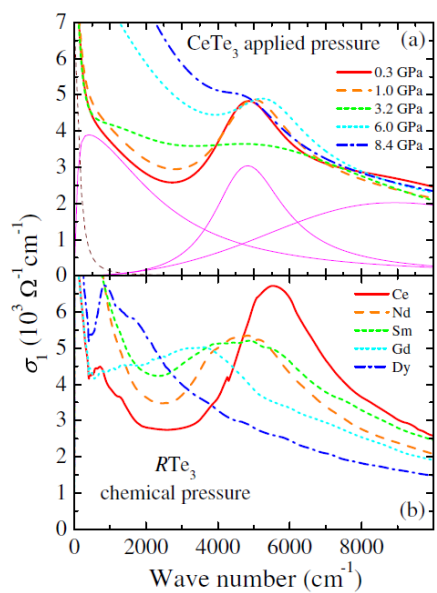}
\end{center}
\caption{
(Color online) 
Optical conductivity ($\sigma_1$) of $R$Te$_3$ ($R=$~Ce, Nd, Sm, Gd, Dy) at room temperature reported by Sacchetti {\it et al.}~\cite{sacchetti-2007}
(a) Pressure dependence of $\sigma_1$ for CeTe$_3$.
The thin solid curves show the decomposition to Drude and Lorentz components.
(b) Data for different $R$ elements at ambient pressure.
Note that the lattice constant of $R$Te$_3$ becomes smaller from Ce to Dy (chemical pressure effect).     
Figure reprinted with permission from Sacchetti {\it et al.}, Phys. Rev. Lett. {\bf 98}, 026401, 2007.
Copyright 2007 by the American Physical Society.  
}
\label{fig:CDW1}
\end{figure}
Charge density waves (CDWs) in solids have a long history of research.~\cite{gruner}
A CDW state is realized in the presence of a strong nesting of Fermi surface (FS) and a strong coupling of conduction electrons with another degree of freedom such as the crystal lattice.
In such a case, the spatial modulation of the charge density with the nesting vector may appear and an energy gap may open at $E_{\rm F}$.
The energy gap (or a CDW gap) can be observed by optical spectroscopy.
Recently, high-pressure IR-SR studies have been performed to analyze CDW states in $R$Te$_3$ ($R$=rare earth)~\cite{sacchetti-2007,lavagnini-2008a,lavagnini-2009} and Pr-filled skutterudite PrRu$_4$P$_{12}$.~\cite{okamura-2012}
$R$Te$_3$'s have a layered crystal structure with strongly two-dimensional (2D) electronic properties with insulating $R$-Te layers and metallic Te layers.  
At room temperature, they are already deep in the CDW state.
However, since the FS nesting is only along the layers, only a part of the FS is gapped in the CDW state.
This leads to a strong Drude component in their $\sigma(\omega)$, where the CDW gap appears as an excitation peak in the IR range.
As shown in Fig.~\ref{fig:CDW1}, Sacchetti {\it et al.} measured the $\sigma(\omega)$ of CeTe$_3$ under high pressure, and also those of other $R$Te$_3$'s with $R=$~Nd, Sm, Gd, and Dy.~\cite{sacchetti-2007}  
It is seen that, for CeTe$_3$ under pressure, the overall spectral weight increases with pressure and the excitation peak shifts to lower energy.
This is a result of CDW gap becoming smaller with pressure, generating more free carriers and increasing the Drude weight.  
The lattice constant of $R$Te$_3$ becomes smaller in the order of Ce, Nd, Sm, Gd, and Dy owing to the lanthanide contraction, and this acts as a chemical pressure.
The spectral evolutions in the two cases of the external (physical) and chemical pressures correspond to each other remarkably well.
Additional pressure measurements on related compounds were also reported.~\cite{lavagnini-2008a,lavagnini-2009}

\begin{figure}[b]
\begin{center}
\includegraphics[width=0.475\textwidth]{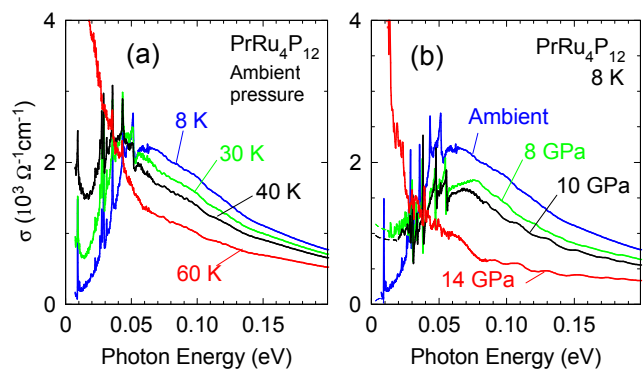}
\end{center}
\caption{
(Color online) 
Optical conductivity ($\sigma$) of the CDW compound PrRu$_4$P$_{12}$ with a transition temperature of 63~K at ambient pressure.
(a) Temperature dependence of $\sigma(\omega)$ at ambient pressure reported by Matsunami {\it et al.}~\cite{matunami-2005}
(b) Pressure dependence of $\sigma(\omega)$ at 8~K reported by Okamura 
{\it et al.}~\cite{okamura-2012}
}
\label{fig:CDW2}
\end{figure}
Another CDW compound, the Pr-filled skutterudite compound PrRu$_4$P$_{12}$, has been studied under high pressure.~\cite{okamura-2012}
Figure~\ref{fig:CDW2} shows the $\sigma(\omega)$ spectra measured at various temperatures and pressures.
At ambient pressure, the CDW transition occurs at 63~K,~\cite{sekine-1997} and the spectra in Fig.~\ref{fig:CDW2}(a) clearly show the opening of a well-developed energy gap with cooling.\cite{matunami-2005}  
The FS nesting in this compound has a 3D character, and the entire FS is gapped in the CDW state, in contrast to the situation of $R$Te$_3$ discussed above.
The pressure-dependent $\sigma(\omega)$ spectra at 8~K in Fig.~\ref{fig:CDW2}(b) show the suppression of energy gap with increasing pressure, which is consistent with the result of earlier electrical resistivity study.~\cite{miyake-2004}
In contrast to the conventional CDW state based on electron-lattice coupling, in PrRu$_4$P$_{12}$, the spatial modulation needed for CDW is provided by the hybridization of Pr~$4f$ electrons with conduction electrons, and by the formation of two Pr sublattices with different schemes of crystal field-split $4f$ levels.~\cite{shiina-2009}
$f$ electron hybridization is usually very weak for Pr compounds, but the Pr ion is surrounded by a cage of 12 phosphorus ligands in the filled skutterudite structure.
This large coordination number enhances such hybridization, resulting in the CDW state.
Therefore, the pressure suppression of CDW in PrRu$_4$P$_{12}$ is mainly due to the pressure tuning of the crystal field and the hybridization.

High-pressure IR-SR studies of low-dimensional electronic systems showing MIT have been reported by Kuntscher {\it et al.}
Titanium oxyhalides TiO$X$ ($X$=Cl, Br) have a spin-Peierls ground state and are Mott-Hubbard insulators with an energy gap.
The pressure evolution of the anisotropic electronic structures of these compounds was probed by polarization-resolved IR-SR spectroscopy.~\cite{kuntscher-2007a,kuntscher-2008}
A gap closure was observed for both compounds at pressures above 10~GPa, and it was coincident with the structural phase transition observed by X-ray diffraction analysis under high pressure.
A quasi-1D conductor LaTiO$_{3.41}$, which is conducting along the $a$-axis, was also studied up to 20~GPa at room temperature.~\cite{kuntscher-2006b} 
In the $\sigma(\omega)$ measured along the $a$-axis, a pronounced mid-IR peak was observed, and it showed marked pressure dependences.
These results were discussed in terms of polarons.  
$\beta$-vanadium bronze ($\beta$-Na$_{0.33}$V$_2$O$_5$), which is a pressure-induced, one-dimensional superconductor with $T_c=$~8~K above 8~GPa, was also studied by IR-SR spectroscopy up to 20~GPa.\cite{kuntscher-2005,kuntscher-2007b}

\subsection{Organic compounds under high Pressure}

Organic metals and semiconductors exhibit a rich variety of physical phenomena under high pressure.
For example, the IR reflectance of the pressure-induced superconductor (TMTSF)$_2$PF$_6$ and the charge-ordering insulator (TMTTF)$_2$PF$_6$ under high pressure have been studied using DAC and IR-SR.~\cite{pashkin-2010}
%
For 2D organic conductor $\kappa$-(ET)$_2$Cu[N(CN)$_2$]$X$ ($X=$~Br, Cl), detailed infrared spectroscopic imaging studies have been performed under high pressure.
This topic is discussed in \S\ref{sec:imaging} in detail.  

\subsection{Elemental metals (Fe, Yb) under high pressure} 

It is well known that many elemental materials exhibit interesting properties under high pressure.  For example, many of them that are not superconducting at ambient pressure become superconducting under high pressure.\cite{shimizu-2005}  Si, which is one of the most typical semiconductors with an energy gap, becomes a metal at pressures above 13~GPa.\cite{syassen-1988}   In addition, the electronic structures of Fe~\cite{seagle-2009} and Yb~\cite{okamura-2007} have been studied under high pressure using IR-SR.   

Elemental Fe undergoes a structural phase transition from a bcc structure (called $\alpha$-Fe) to a hcp structure ($\varepsilon$-Fe) at 13--18~GPa~\cite{wang-1998} and changes in the electronic structure upon this transition are of interest from physical and geophysical points of view.  The $R(\omega)$ of elemental Fe at room temperature was measured by Seagle {\it et al.} in the 1000--8000~cm$^{-1}$ range under external pressures up to 50~GPa.~\cite{seagle-2009}  The measured $R(\omega)$ was nearly constant below 13~GPa, but showed large changes in its slope from 13 to 20~GPa, which is clearly due to the bcc-hcp transition.  Above 20~GPa, although the structure remained the same, $R(\omega)$ kept decreasing with pressure.  At 50~GPa, $R(\omega)$ at 5000~cm$^{-1}$ was about 60\% of that at ambient pressure.

Elemental Yb also shows unique properties under pressure.  Although most elemental rare earths are trivalent metals at ambient pressure, Yb is divalent with a filled $4f$ shell.  It undergoes a structural phase transition at 4~GPa from an fcc structure to a bcc structure, although the latter has a lower volume filling.  The electrical resistivity of Yb increases with pressure up to the fcc-bcc transition.~\cite{mcwhan-1969}  The Yb valence increases with pressure, from 2 at ambient pressure to about 2.6 at 20~GPa.~\cite{syassen-1982a,fuse-2004}  The $R(\omega)$ of elemental Yb at room temperature in the 0.03--1.1~eV range was measured under high pressure up to 18~GPa by Okamura {\it et al.}~\cite{okamura-2007}  $R(\omega)$ developed a dip at approximately 0.1~eV with increasing pressure up to 4~GPa.  This suggested a decrease in the density of states near $E_{\rm F}$, which corresponds well to an increase in resistivity.  Above 4~GPa, the above dip disappeared, but the overall IR $R(\omega)$ gradually decreased with further pressure, showing a deep ``valley'' centered at 0.2~eV.  They speculated that the increase in the Yb valence, namely, the involvement of $4f$ electron state near $E_{\rm F}$, might be related with the development of this feature.

\subsection{Rare-earth (``heavy-fermion'') compounds}

Another important group of materials belonging to SCES are rare-earth-based intermetallic compounds containing $f$ electrons.~\cite{onuki-2004}
$f$ electrons in these compounds are basically localized in the $4f$ orbital located in the inner shell.  However, their hybridization with conduction ($c$) electrons may lead to a partially delocalized state at low temperatures, with an increased effective mass ($m^\ast$).  Large $m^\ast$ values are manifested in, for example, the linear temperature coefficient ($\gamma$) of the electronic specific heat.  $\gamma$ values of up to 1000~mJ/K$^2$mol and above have been observed, in contrast to $\sim$~1~mJ/K$^2$mol usually observed for simple metals.  The hybridization between the conduction and $4f$ electrons, namely, $c$-$f$ hybridization, mass enhancement, and the duality between localized and delocalized states are central issues in the physics of $f$ electron systems.~\cite{onuki-2004}  Most representative systems are Ce- and Yb-based compounds.  Trivalent Ce has a $4f^1$ configuration, and trivalent Yb has a $4f^{13}$ configuration with a hole in the $f$ orbital.  When the hybridization is strong, the average $f$ occupancy decreases for Ce, and increases for Yb; hence the average valence increases and decreases for Ce and Yb, respectively.  These ``intermediate-valence'' (IV) states are generally characterized with the paramagnetic ground state with a moderately increased $m^\ast$, in the range from ten to several hundreds of mJ/K$^2$mol.\cite{lawrence-2008}  For more localized systems, a transition to a magnetically ordered, typically antiferromagnetic (AFM), state is often observed.~\cite{onuki-2004}  Such compounds generally have larger $\gamma$ values than the IV systems.  In addition, a quantum critical transition may be induced from the magnetically ordered state by chemical doping, and by applying an external pressure or a magnetic field.  In the vicinity of such a quantum critical point (QCP), various anomalous physical properties have been observed, such as unconventional superconductivity.~\cite{sarrao-2007}

Measurements of $\sigma(\omega)$ spectra have been very useful for probing unique peculiar electronic structures of $f$ electron systems.~\cite{Degiorgi1999,Basov2011}
For example, $\sigma(\omega)$ spectra show an extremely narrow Drude peak due to the dynamics of heavy quasiparticles.~\cite{Degiorgi1999}
In addition, a characteristic mid-IR peak is often observed in $\sigma(\omega)$, which results from a $c$-$f$ hybridized state far from $E_{\rm F}$.~\cite{hancock-2000,dordevic-2001,degiorgi-2001,Kimura2006b,okamura-2007b,kimura-2009a,kimura-2009b,Iizuka2010,kimura-2011a,kimura-2011b,kimura-2011c}
So far, high pressure IR-SR studies of the $f$ electron state have been reported for SmTe~\cite{Kwon1995}, SmS~\cite{mizuno-2008}, YbS~\cite{matunami-2009}, CeRu$_4$Sb$_{12}$~\cite{okamura-2011}, and CeIn$_3$~\cite{Iizuka2012}.
Both SmS and YbS are divalent semiconductors at ambient pressure, but undergo pressure-induced transition to a metallic IV state.
CeRu$_4$Sb$_{12}$ is a Ce-filled skutterudite compound, and is an IV compound with a moderate mass enhancement.
Under pressure, its electrical resistivity increases, and the measured $\sigma(\omega)$ spectra suggest a pressure-induced crossover to a semiconducting state at low temperatures.~\cite{okamura-2012}
CeIn$_3$ has more localized $f$ electron characters, with an AFM ground state at ambient pressure.
Below, high-pressure IR-SR studies of YbS and CeIn$_3$ are reviewed.   

\subsubsection{Pressure Tuning of an Ionic Semiconductor into a Heavy Electron Metal in YbS}

YbS at ambient pressure is an ionic (Yb$^{2+}$S$^{2-}$) insulator (semiconductor) with a rocksalt structure.  Its energy gap is about 1.3~eV, and is located between a fully occupied $4f$ state and an unoccupied, mainly Yb $5d$-derived conduction ($c$) band.  An optical reflectance study of YbS using DAC was performed in 1980s by Syassen {\it et al.}, without using IR-SR.~\cite{syassen-1985}  A peak in $\sigma(\omega)$ due to light absorption across the energy gap, namely, a gap excitation peak, was observed to shift to a lower energy with increasing pressure.  The peak shift was extrapolated to zero energy at about 10~GPa, implying a gap closure near 10~GPa.  Above 10~GPa, a pronounced near-IR peak was observed in the $\sigma(\omega)$ spectrum.   In addition, an increase in mean Yb valence from 2, reaching about 2.4 at 20~GPa, was suggested by the volume compression data of YbS.\cite{syassen-1985}  From these data, an IV metallic state above 10~GPa was implied.  However, their work was performed at photon energies above 0.5~eV, and detailed information about lower-energy electronic structures was not obtained.

\begin{figure}[b]
\begin{center}
\includegraphics[width=0.4\textwidth]{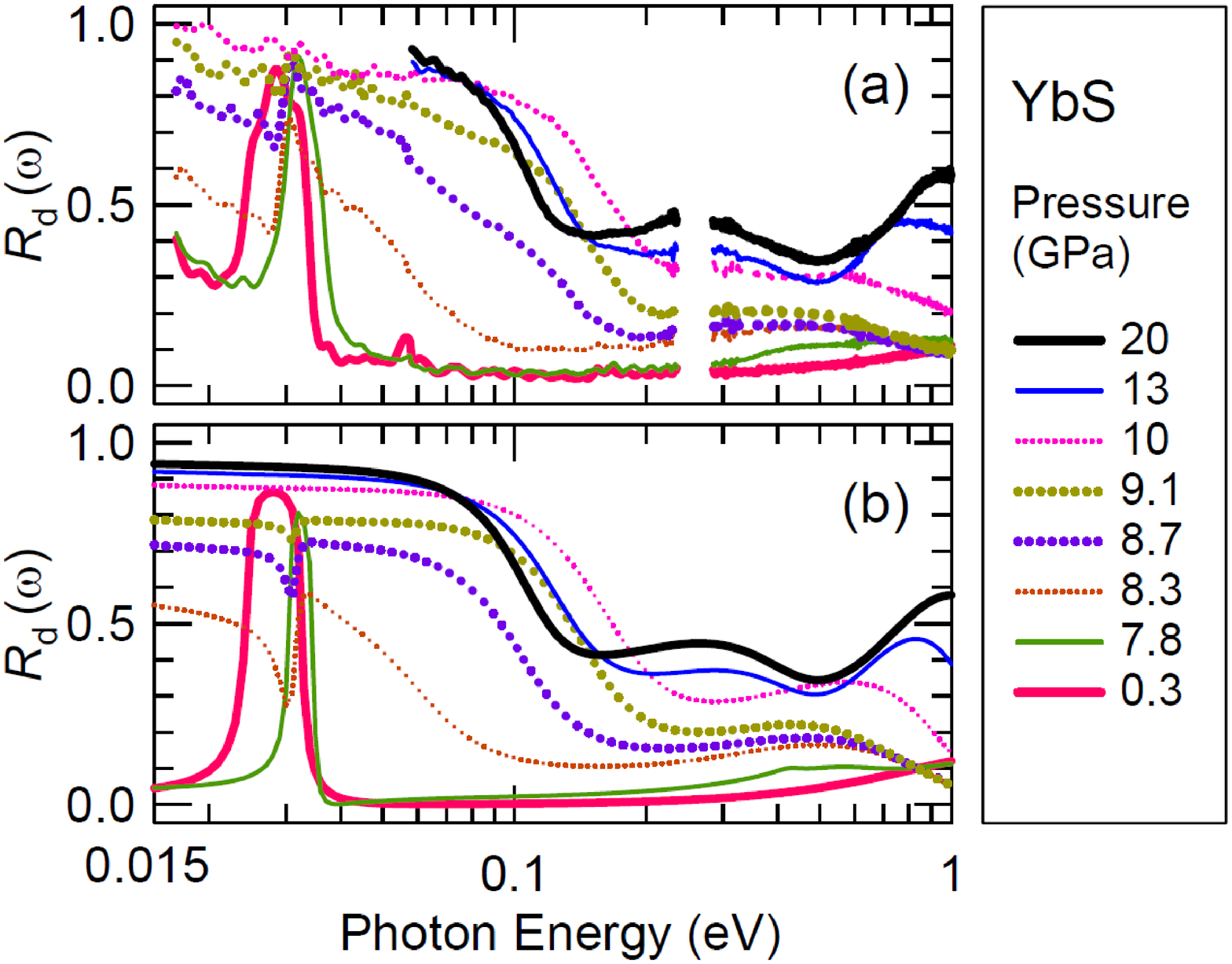}
\end{center}
\caption{
(Color online) 
Optical reflectance [$R_d(\omega)$] of YbS under external pressure of up to 20~GPa, reported by Matsunami {\it et al.}~\cite{matunami-2009}  
(a) Data measured in DAC at room temperature.  (b) Results of Drude-Lorentz fitting to the data in (a), performed as described in \S\ref{sec:DL}.  
}
\label{fig:ybs-R}
\end{figure}
Matsunami {\it et al.} extended the high-pressure IR study of YbS down to 20~meV using IR-SR.~\cite{matunami-2009}  Figure~\ref{fig:ybs-R}(a) shows their reflectance [$R_d(\omega)$] spectra of YbS measured with DAC over the 0.02--1.1~eV energy range and under pressures up to 20~GPa.  The missing portion between 0.2 and 0.3~eV is due to strong absorption by the diamond anvil.  At 0.3~GPa, the $R_d(\omega)$ spectrum is low except for the phonon peak centered near 0.03~eV.  The $R_d(\omega)$ spectrum does not change very much up to 7.8~GPa, but then shows a marked increase with pressure.  Namely, $R_d(\omega)$ spectrum below 0.1~eV shows a rapid increase with pressure, reaching about 0.9 above 10~GPa, and the phonon peak disappears quickly with increasing pressure.  The high-reflectance band is clearly due to plasma reflection.  These data indicate that the energy gap closes at about 8~GPa, and that YbS above 10~GPa becomes a metal.  In addition to the plasma reflection, two marked dips are observed to develop in the $R_d(\omega)$ spectra above 10~GPa.  Figure~\ref{fig:ybs-R}(b) shows the results of the Drude-Lorentz fitting to the measured $R_d(\omega)$ spectra in Fig.~\ref{fig:ybs-R}(a).  The fitting was performed as described in \S\ref{sec:DL}, including the refractive index of diamond through eq.~(\ref{eq:fresnel2}).  The fitted spectra in Fig.~\ref{fig:ybs-R}(b) well reproduce the overall pressure evolution of $R_d(\omega)$ in Fig.~\ref{fig:ybs-R}(a).

\begin{figure}[b]
\begin{center}
\includegraphics[width=0.4\textwidth]{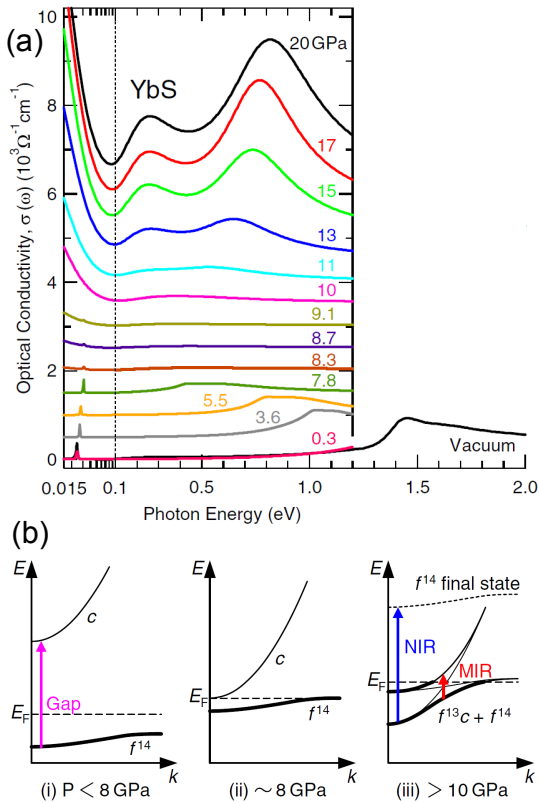}
\end{center}
\caption{
(Color online) 
Optical conductivity $\sigma(\omega)$ spectra of YbS under external pressures up to 20~GPa reported by Matsunami {\it et al.}~\cite{matunami-2009}
(a) $\sigma(\omega)$ spectra obtained from the Drude-Lorentz fitting of $R_d(\omega)$, shown in Fig.~\ref{fig:ybs-R}.
$\sigma(\omega)$ measured in vacuum is also indicated.
The peak due to the absorption edge is seen to shift to lower energies with pressure.
Above 8~GPa, a narrow Drude peak appears, and two peaks also appear in the mid-IR and near-IR ranges.
(b) Proposed model of the microscopic electronic structures of YbS.
}
\label{fig:ybs-s}
\end{figure}
Figure~\ref{fig:ybs-s}(a) shows the $\sigma(\omega)$ spectra given by the fitting, together with the $\sigma(\omega)$ measured in vacuum without using DAC and using the usual KK analysis.   With increasing pressure, the gap excitation peak originally located at 1.3~eV in the vacuum data is seen to shift to a lower energy, becomes weaker, and almost disappears at 8.3~GPa.  Corresponding to the appearance of plasma reflection in $R_d(\omega)$, a Drude peak starts to increase in $\sigma(\omega)$ above 8~GPa, which is very narrow.  Above 10~GPa, furthermore, two pronounced peaks develop near 0.25 and 0.7--0.8~eV.  
Hereafter, these peaks are referred to as the mid-IR peak and near-IR peak, respectively.
These peaks strongly suggest that the metallic phase above 10~GPa is not a simple metal, but involves strong structures in its density of states near $E_{\rm F}$.
Figure~\ref{fig:ybs-s}(b) shows the proposed model for the pressure evolution of the microscopic electronic structure in YbS: 
Under low pressure, as shown in Fig.~\ref{fig:ybs-s}(b)(i), YbS is an insulator with an energy gap between the Yb $5d$ $c$ band and the fully occupied Yb $4f$ state.(The S 3$p$ valence band is located about 4~eV below $E_{\rm F}$.)
An LDA+$U$ band calculation of YbS at ambient pressure shows that the $c$ band bottom is located at the $X$ point in the Brillouin zone, and at the top of the $f$ band at the $\Gamma$ point.~\cite{harima-2011}
With increasing pressure, the bandwidths of both the $c$ and $f$ bands increase owing to the reduced interatomic distance.
This should reduce the energy gap, resulting in the observed red shift of the gap excitation peak in $\sigma(\omega)$.
At around 8~GPa, as shown in Fig.~\ref{fig:ybs-s}(b)(ii), the $c$ and $f$ bands start to overlap, and the energy gap closes.
Hence, the gap excitation peak disappears and the Drude peak starts to grow.
Above 10~GPa, as shown in Fig.~\ref{fig:ybs-s}(b)(iii), the overlap between the $c$ and $f$ bands becomes larger, and the hybridization between the two bands may lead to an anticrossing behavior.
This may open up a new channel for interband transition and may be responsible for the appearance of a mid-IR peak above 10~GPa.
In addition, this hybridized state is necessarily an IV state, since $c$ electrons are transferred from the $f$ band to the $c$ band, leaving behind holes in the $f$ band.
This is consistent with the observed increase in the mean Yb valence above 10~GPa.~\cite{syassen-1985}
Assuming that the increase in the mean Yb valence from 2 is equal to the density of $c$ electrons, the analysis of spectral weight for the Drude peak indicated that the effective mass of $c$ electrons is about 12~$m^\ast$ at 20~GPa.
Namely, carriers in the metallic phase are moderately heavy.
The origin for the near-IR peak is unclear, but it may be a result of excitation from the $f^{13}c$ component of the hybridized state back to the $f^{14}$ state.

\subsubsection{Continuity of electronic structure of CeIn$_3$ across critical pressure}

As introduced before, physics at QCP, which is the border between local magnetism and itinerant paramagnetism at zero temperature, has become one of the main topics in the condensed-matter field because new quantum properties such as non-BCS superconductivity appear in the vicinity of the QCP. 
The ground state of rare-earth heavy-fermion compounds changes between the local magnetic and itinerant nonmagnetic states through external perturbation by such factors as pressure and magnetic field~\cite{Gegenwart2008}. 
The QCP appears owing to the energy balance between the local magnetic state based on the Ruderman-Kittel-Kasuya-Yoshida (RKKY) interaction and the itinerant heavy-fermion state due to the Kondo effect.~\cite{onuki-2004}
In the itinerant heavy-fermion regime, the conduction band hybridizes with the nearly local $4f$ state, so that a large Fermi surface, as well as the hybridization band between them, namely, the $c$-$f$ hybridization band, is realized. 
In the case of a magnetic regime, on the other hand, two theoretical scenarios have been proposed. 
One is the spin-density wave (SDW) scenario based on spin fluctuation, in which large Fermi surfaces due to $c$-$f$ hybridization remain even in magnetically ordered states. 
The other is the Kondo breakdown (KBD) scenario, in which the $c$-$f$ hybridization state disappears in the magnetic state and only small Fermi surfaces due to conduction electrons appear~\cite{Coleman2001}.
Recently, Iizuka {\it et al.} have studied the $c$-$f$ hybridization state of CeIn$_3$ under high pressure by far-IR $R(\omega)$ measurement~\cite{Iizuka2012}. 
Then the SDW scenario was concluded to be more suitable for the electronic structure of CeIn$_3$ than the KBD scenario.

Optical measurements are highly useful means of clarifying the electronic structure at a critical pressure.
If such optical measurements are performed from the local regime to the itinerant regime with varying pressure or magnetic field, changes in the electronic structure across the QCP may be revealed.  
One candidate material that can be studied across a critical pressure at an accessible temperature above 5~K is CeIn$_3$.  
CeIn$_3$ has an AFM ground state with a N\'eel temperature $T_{\rm N}$ of 10 K. 
With the application of pressure, $T_{\rm N}$ monotonically decreases and disappears at a critical pressure of approximately 2.6 GPa~\cite{Mathur1998,Knebel2001,Grosche2001}.

\begin{figure}[b]
\begin{center}
\includegraphics[width=0.45\textwidth]{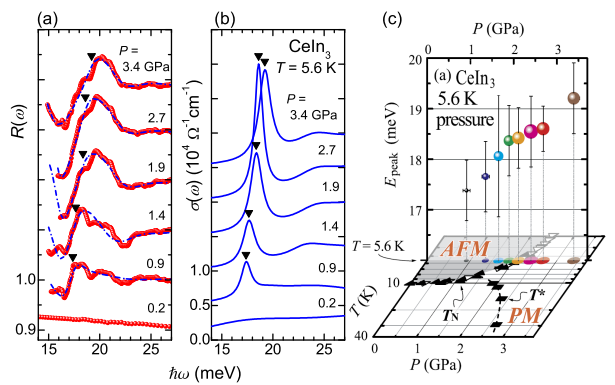}
\end{center}
\caption{
(Color online)
(a) Pressure dependence of the reflectivity [$R(\omega)$] spectrum (solid circles) of CeIn$_3$ and DL fitting results (dotted-dashed lines) in the photon energy $\hbar\omega$ range of 14--27~meV at 5.6~K. 
The spectra are shifted by 0.1 for clarity.   
(b) Optical conductivity [$\sigma(\omega)$] spectra derived from the DL fitting of the $R(\omega)$ spectra in (a).  The Drude component was subtracted.  These spectra are offset by $5\times10^3 \Omega^{-1}{\rm cm}^{-1}$ for clarity.  The solid triangles show the center energies of the fitted Lorentzian functions.  (c) Edges of $R(\omega)$ in (a) and peak energies ($E_{peak}$) of $\sigma(\omega)$ in (b) as a function of pressure at 5.6~K.   
The size of the marks denotes the intensity of the corresponding peak in the $\sigma(\omega)$ spectra. 
The pressure-dependent N\'eel temperature (solid and open triangles, $T_{\rm N}$) and valence transition temperature (solid squares, $T^*$)~\cite{Kawasaki2001} are also plotted at the bottom.
From Iizuka {\it et al.}, 2012.~\cite{Iizuka2012}
}
\label{Pdep}
\end{figure}
To investigate the pressure effect of the $c$-$f$ hybridization gap, pressure-dependent reflectivity [$R(\omega, P)$] spectra were measured at 5.6~K as shown in Fig.~\ref{Pdep}(a).
The $R(\omega,P = 0.2~{\rm GPa}$) spectrum is almost flat, because the absorption in the $c$--$f$ hybridization band is not large.
With increasing pressure, however, a significant dispersive structure appears and its intensity increases.
To obtain the pressure-dependent optical conductivity [$\sigma(\omega,P)$] spectra from the $R(\omega,P)$ spectra, DL fitting as shown in \S\ref{sec:DL} was applied~\cite{DresselGruner}.  The parameters for the Drude part were fixed, because the change in the Drude weight cannot be recognized from the spectra owing to the limited spectral region.  
The obtained $\sigma(\omega,P)$ spectra are shown in Fig.~\ref{Pdep}(b), where a pronounced peak is observed to grow with increasing pressure.  This peak corresponds to the dispersive structure in $R(\omega, P)$ mentioned above, 
and it originates from a $c$--$f$ hybridization gap formed under pressure.  

The energy and the effective electron number of the $\sigma(\omega)$ peaks in Fig.~\ref{Pdep}(b) are plotted using the position and size of marks, respectively, as a function of pressure in Fig.~\ref{Pdep}(c). 
The figure shows that the edge of $R(\omega)$ as well as the $\sigma(\omega)$ peak monotonically shifts to the higher-energy side and grows with increasing pressure.   These results indicate that the $c$-$f$ hybridization state appears even in the AFM phase.  
In Fig.~\ref{Pdep}(c), the pressure-dependent $T_{\rm N}$ and the crossover temperature $T^*$ between the localized and itinerant regimes of the $4f$ electrons observed in a previous nuclear quadrupole resonance (NQR) experiment are also plotted at the bottom~\cite{Kawasaki2001}.
The observed emergence pressure ($\sim$1.6~GPa) of the $c$-$f$ hybridization gap at 5.6~K is roughly located on the extended line of $T^*$.
Therefore, the $T^*$ line can be extended from the PM phase to the AFM phase, further verifying the applicability of the SDW scenario.~\cite{Gegenwart2008}

\subsection{High-pressure IR studies without using IR-SR}\label{sec:no-SR}

We have emphasized the advantage of using high-brilliance IR-SR for high pressure experiments, but many high pressure IR experiments have also been performed without IR-SR.
They are briefly reviewed here.

Before IR-SR became available, high pressure optical studies had been performed with DAC and conventional light sources.
For example, in the early 1980s, Syassen and Sonnenschein developed a high-pressure optical spectroscopy apparatus that covered the near IR to UV ranges (0.5--4~eV).~\cite{syassen-1982b}
Syassen and coworkers studied various materials under high pressure such as K (potassium),~\cite{syassen-1983} EuO,~\cite{syassen-1984} YbS and YbO,~\cite{syassen-1985} and Si.~\cite{syassen-1988}
At present, commercially available FTIR instruments coupled with a microscope (micro-FTIRs) generally cover a wide spectral range with a high spatial resolution and a high signal-to-noise ratio.~\cite{griffiths-2007,griffiths-2009}
This has made it much easier to work with a DAC.
For example, Pashkin {\it et al.}, without using IR-SR, measured the $R(\omega)$ of organic compound (TMTTF)$_2$AsF$_6$ under pressure of up to 6~GPa over 550--8000~cm$^{-1}$ range.~\cite{pashkin-2006}
In addition, high-pressure reflectance studies have been performed down to 6~meV (50~cm$^{-1})$ for elemental Bi,~\cite{armitage-2010} and down to 4~meV (30~cm$^{-1}$) for TaS$_2$,~\cite{kezsmarki-2007} without using SR, although the pressure was limited to below 2.5~GPa owing to the large diamond required to cover the long-wavelength range.
Mita {\it et al.} studied the $R(\omega)$ of the Mott insulator MnO up to 140~GPa without IR-SR and observed metallization at 94~GPa, although they used powder samples and the spectral range was limited to above 2000~cm$^{-1}$.~\cite{mita-2005}

However, if one wants to perform high-pressure study in the far-IR/THz range and to go up to several GPa and above, it would be difficult with a thermal source owing to its low brightness.  The ability to reach several 5-7~GPa or above is quite advantageous in working with SCES, since many interesting physical properties have been found in that pressure range.  Even in the mid-IR, the higher brightness of IR-SR may enable experiments under more difficult conditions.  
For example, higher pressures (i.e., smaller diamond anvils and therefore smaller samples) may be obtained with SR.  In addition, studies of crystal samples rather than of powder ones may become possible.  The ability to work with a crystal sample is quite important in studying SCES, since grinding the sample into powders may alter the optical properties themselves, since even the polishing of a large single crystal sometimes alters the optical properties for certain materials.~\cite{okamura-2007c}

\section{IR/THz Spectroscopy under Magnetic Field}\label{sec:mag}

\begin{figure}[b]
\begin{center}
\includegraphics[width=0.45\textwidth]{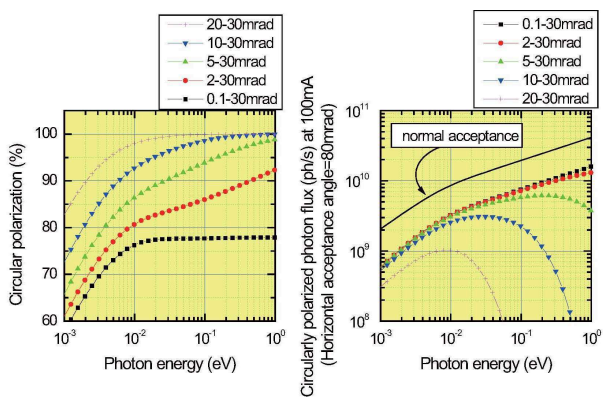}
\end{center}
\caption{
(Color online)
Calculated circular polarization and circularly polarized photon flux from the IR/THz beamline BL6B of UVSOR-II with several acceptance angles.
}
\label{CP}
\end{figure}
The high brightness and low emittance of IR/THz-SR are also advantageous for spectroscopic studies under a high magnetic field using a superconducting magnet, because the sample must be set in the restricted space at the center of a long and narrow magnet bore.
To produce a magnetic field of 10~T or more, the room-temperature bore radius of the superconducting magnet must be less than 30~mm, and focusing and collecting mirrors must be set about 500~mm or farther from the center of the magnet.
When a normal-incident reflectivity measurement is performed, the required emittance is less than 30~$\mu$m$\cdot$rad if the sample diameter is about 1~mm because the incident and reflected light beams must be carried through the bore. 
Therefore, a thermal IR source is not suitable for such a higher magnetic field because the emittance of a conventional IR source is $\sigma\cdot\sigma'$~=~66~$\mu$m$\cdot$rad.\cite{footnote}
For example, optical studies with (unpolarized) IR-SR and a superconducting magnet have been reported on the electronic structures of the colossal magnetoresistance (CMR) material Tl$_2$Mn$_2$O$_7$~\cite{okamura-2001} and a superconducting Nb$_{0.5}$Ti$_{0.5}$N film.~\cite{tanner-2010}   
The electron spin resonance has also been measured for various quantum spin systems such as Ni$_5$(TeO$_3$)$_4$Cl$_2$.~\cite{mihaly-2006}

In addition to the high brilliance and low emittance, the linear/circular polarization of IR-SR can be very useful for high-field studies.
Since SR is emitted from the electron beam in an accelerator, the emitted light has a linear polarization in the orbital plane of the electron beam.
It also has circular or elliptical polarization off the orbital plane, which is a projection of the electron beam motion.
As an example, circular polarization that can be obtained at UVSOR-II is shown in Fig.~\ref{CP}.~\cite{Kimura1999}
It is seen that a circular polarization is obtained over a very wide spectral range, without additional optical elements.
In contrast, when a conventional, thermal IR source is used to produce linear or circular polarization, optical elements such as linear polarizers and quarter wavelength ($\lambda/4$) plates are required.
These optical elements limit the spectral range and therefore the experimental ability.
With IR-SR, one can use a much wider spectral range of linearly and circularly polarized light.
So far, experiments using circular polarized IR-SR under a magnetic field have been performed to study the magnetic excitons in GdAs~\cite{Kimura1998}, the magneto-plasma resonance in Tl$_2$Mn$_2$O$_7$~\cite{okamura-2005}, and the electronic structures of CeSb and CeBi in complex magnetic phases~\cite{Kimura2000-1,Kimura2000-2,Kimura2002}.
Below, the results of Tl$_2$Mn$_2$O$_7$, Nb$_{0.5}$Ti$_{0.5}$N, and CeSb are reviewed.

\subsection{Magnetic-field-dependent IR spectroscopy}

\begin{figure}[b]
\begin{center}
\includegraphics[width=0.475\textwidth]{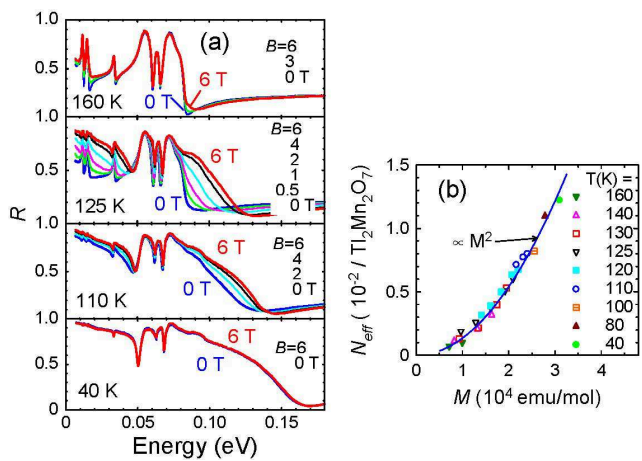}
\end{center}
\caption{
(Color online) 
IR spectroscopy of CMR compound Tl$_2$Mn$_2$O$_7$ reported by Okamura {\it et al.}~\cite{okamura-2001}
(a) Reflectances ($R$) of Tl$_2$Mn$_2$O$_7$ under magnetic field ($B$) at four temperatures.
(b) The effective carrier density $N^\ast$ derived from the reflectance data plotted as a function of $M^2$, where $M$ is the magnetization of the sample.
}
\label{fig:TMO}
\end{figure}
Tl$_2$Mn$_2$O$_7$ with the pyrochlore structure shows a CMR near its Curie temperature, $T_{\rm C}=$~120~K, where the resistivity decreases by an order of magnitude in a magnetic field of 7~T.\cite{shimakawa-1996}  
The tetravalent Mn$^{4+}$ is Jahn-Teller-inactive; hence, the mechanism of CMR is different from that of perovskite manganites such as La$_{1-x}$Sr$_x$MnO$_3$ with Mn$^{3+}$.\cite{tokura-1994}   
In Tl$_2$Mn$_2$O$_7$, conduction occurs in the Tl~$6s$--O~$2p$ conduction band, while a ferromagnetic order occurs in the Mn sublattice independently of the conduction channel.
Figure~\ref{fig:TMO}(a) shows the measured $R(\omega)$ spectra under magnetic fields ($B$).\cite{okamura-2001}   
At 125~K, $R(\omega)$ has a typical spectral shape of an insulator at $B=0$, but it remarkably increases with $B$, showing a clear Drude component at $B=6$~T.
At other temperatures far from $T_{\rm C}$, however, $R(\omega)$ show only minor changes.
Figure~\ref{fig:TMO}(b) shows a plot of the effective carrier density ($N_{eff}$), which was calculated from the $\sigma(\omega)$ spectra obtained from $R(\omega)$, as a function of the magnetization ($M$) at various temperatures and magnetic fields, which was measured on the same sample used for the $R(\omega)$ study.  
$N_{eff}$ is well scaled with $M^2$ over wide ranges of temperatures and $B$ fields.
These results demonstrate that the carrier density of Tl$_2$Mn$_2$O$_7$ is directly related with magnetization, rather than to the external magnetic field or temperature.
The microscopic mechanism for this $M^2$ scaling is unclear. 
However, it is remarkable that a very similar scaling between $N_{eff}$ and $M^2$ has also been observed for another magnetoresistive material EuB$_6.$\cite{degiorgi-1997}

\begin{figure}[b]
\begin{center}
\includegraphics[width=0.4\textwidth]{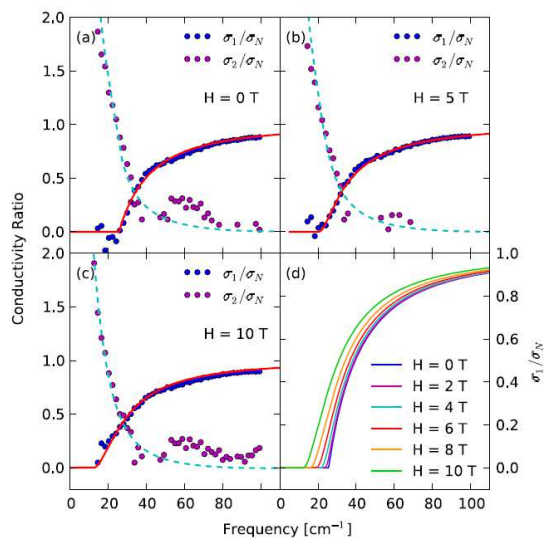}
\end{center}
\caption{
(Color online) 
Real part ($\sigma_1$) and imaginary part ($\sigma_2$) of the complex optical conductivity measured on a Nb$_{0.5}$Ti$_{0.5}$N film reported by Xi {\it et al.}~\cite{tanner-2010}
The data were measured in the superconducting state at 3~K and normalized by that in the normal state at 10~K.
The dots are the data, and the solid curves are fit using a microscopic theory.
The curves in (d) clearly show the reduction in superconducting gap with magnetic field.   
Figure reprinted with permission from Xi {\it et al.}, Phys. Rev. Lett. {\bf 105}, 257006, 2010.
Copyright 2010 by the American Physical Society.  
}
\label{fig:SC-mag}
\end{figure}
The magnetic field suppression of the microscopic superconducting state, namely, the breaking of Cooper pairs and the reduction in superconducting gap, has been studied by Xi {\it et al.}, using thin film samples of Nb$_{0.5}$Ti$_{0.5}$N under magnetic fields of up to 10~T.~\cite{tanner-2010}
They studied both the $R(\omega)$ and $T(\omega)$ (transmittance) of a film sample to obtain its $\sigma(\omega)$ at 10--100~cm$^{-1}$ (1.2--12~meV) range.
Figure~\ref{fig:SC-mag} shows the measured $\sigma(\omega)$ at 0, 5, and 10~T.
In a superconducting state with an energy gap of 2$\Delta$, photons having energies less than 2$\Delta$ cannot break up Cooper pairs.  They therefore 
cannot cause electronic excitations, and are reflected without losing energy.  
Accordingly, $\sigma(\omega)$ decreases to zero at photon energies below 2$\Delta$.  
This is clearly demonstrated in the data of Fig.~\ref{fig:SC-mag}, and the fit using the microscopic theory of Cooper pair breaking (solid curves in Fig.~\ref{fig:SC-mag}) shows the reduction in superconducting gap with magnetic field, as shown in Fig.~\ref{fig:SC-mag}(d).
From the data, the pair breaking parameter was also derived as a function of $B$.

\subsection{IR magnetic-circular dichroism of CeSb}

\begin{figure}[b]
\begin{center}
\includegraphics[width=0.45\textwidth]{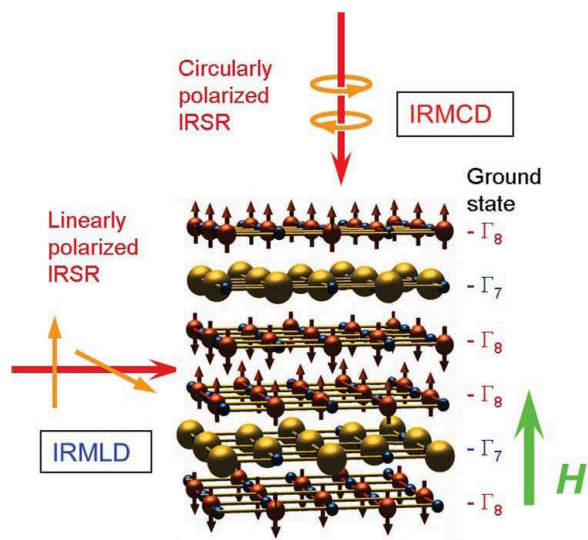}
\end{center}
\caption{
(Color online)
Configuration of the IRMCD and IRMLD measurements of CeSb in the AFP1 phase (See Fig.~\ref{CeSb_R}b) of the magnetic structure of $+0-$.
Balls with arrows indicate Ce$^{3+}$ ions of the $\Gamma_8$ ground state with a magnetic moment of $2\mu_{\rm B}$, large balls paramagnetic Ce$^{3+}$ ions of the $\Gamma_7$ ground state, and small balls Sb$^{3-}$ ions.
From Kimura {\it et al.}, 2002.~\cite{Kimura2002}
}
\label{CeSbAFP1}
\end{figure}
CeSb is a typical SCES with a low carrier concentration.
Despite the simple NaCl-type crystal structure, it has a highly complex magnetic phase diagram with 16 phases.~\cite{RM1985}
Because of the complex magnetic phase diagram and its heavy-Fermion-like physical properties, the material has attracted much attention over the past few decades. 
The origin of the complex magnetic phase diagram was qualitatively explained by the Ce~$4f$--Sb~$5p$ mixing model ($p$-$f$ mixing).~\cite{TakahashiKasuya1985}
The model succeeded in explaining the heavy Fermi surface named $\beta_4$ in the ferromagnetic phase in the de Haas-van Alphen experiments.~\cite{Kitazawa1988}
In magnetically ordered states, double-layer structures of magnetic moments appear.
For instance, the magnetic structure of the antiferromagnetic (AF-1A) phase is $+ + - -$. 
Here,$+$ ($-$) indicates that the magnetic moment is parallel (antiparallel) to the magnetic field. 
The $p$-$f$ mixing model cannot completely explain the double-layer structure. 
The electronic structure near $E_{\rm F}$ directly reflects the change due to the $p$-$f$ mixing and other mixing effects, particularly the Sb~$5p$--Ce~$5d$ mixing ($p$-$d$ mixing) effect.
Therefore, the investigation of the electronic structure in magnetically ordered states gives us useful information on the interactions for the magnetically ordered states.
To clarify the change in the electronic structure in magnetic fields, magnetic-field- and temperature-dependent $R(\omega)$ and $\sigma(\omega)$ spectra were measured, i.e., the spectral change in the magnetic field--temperature ($H-T$) phase diagram.
For a detailed investigation of the spin polarized electronic structure in ordered phases, IRMCD and IR magnetic linear dichroism (IRMLD) experiments were performed.

In the IRMCD experiment, the incident light direction was set parallel to the magnetic field direction (Faraday configuration) and was perpendicular to the (001) plane of CeSb.
The configuration of the plus magnetic field and the left circularly polarized light can excite electrons with a magnetic quantum number ($\Delta m_j$) of $-1$ and, inversely, that with the right circularly polarized light excites electrons with $\Delta m_j=+1$.
On the other hand, in the case of the IRMLD experiment, the incident light direction was set perpendicular to the magnetic field (Voigt configuration).
The IRMLD measurement can detect the anisotropic electronic structure induced by the external magnetic field.
One example of this configuration for the AFP1 phase of CeSb is shown in Fig.~\ref{CeSbAFP1}.~\cite{Kimura2002}

\begin{figure}[b]
\begin{center}
\includegraphics[width=0.45\textwidth]{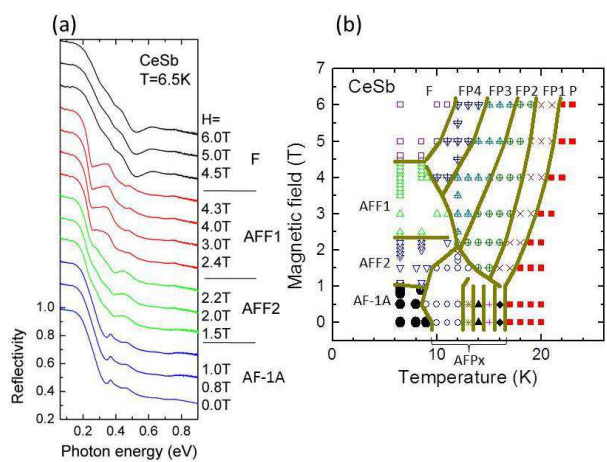}
\end{center}
\caption{
(Color online)
(a) Magnetic field dependence of the reflectivity spectrum of CeSb at 6.5~K using unpolarized light as a function of the photon energy.
Successive curves are offset by 0.16 for clarity.
(b) Temperatures and magnetic fields at which reflectivity spectra of CeSb were measured.
The same marks indicate those at which the same spectra were observed.
The reflectivity spectra were taken with increasing temperature and magnetic field.
The solid lines are where the spectrum changes.
The legends for the magnetic phase follow the same notation as for neutron scattering.~\cite{RM1985}
From Kimura {\it et al.}, 2002.~\cite{Kimura2002}
}
\label{CeSb_R}
\end{figure}
Figure~\ref{CeSb_R}(a) shows the temperature and magnetic field dependences of the $R(\omega)$ spectrum using unpolarized light.
Both of the light and magnetic field directions were set to the [001] direction of the sample.
The $R(\omega)$ spectrum markedly changes with magnetic field and temperature, because the Sb~$5p$ band strongly couples with the magnetic structure of Ce~$4f$ magnetic moments through $p$-$f$ mixing.
Thus, the change in the electronic structure due to magnetic ordering is reflected in the $R(\omega)$ spectra.
In addition, the phase change is monitored using the spectra.
The critical magnetic field and temperature were observed with hysteresis indicating first-order transition.
The temperatures and magnetic fields at which the $R(\omega)$ spectra were measured are plotted in Fig.~\ref{CeSb_R}(b).
The same marks in the figure are those at which exactly the same spectra were observed.
The change in the spectrum is indicated by solid lines.
These lines must be phase boundaries.
According to the magnetic phase diagram derived from neutron scattering,~\cite{RM1985} the boundaries determined by our $R(\omega)$ measurement are in good agreement with the magnetic phase diagram.
Therefore, the change in the $R(\omega)$ spectrum indicates a magnetic phase transition.
In addition, the $\sigma(\omega)$ spectra reflect the electronic structures in magnetically ordered states.
Thus, the $\sigma(\omega)$ spectra provide information on the electronic structure in such ordered states.

\begin{figure}[b]
\begin{center}
\includegraphics[width=0.45\textwidth]{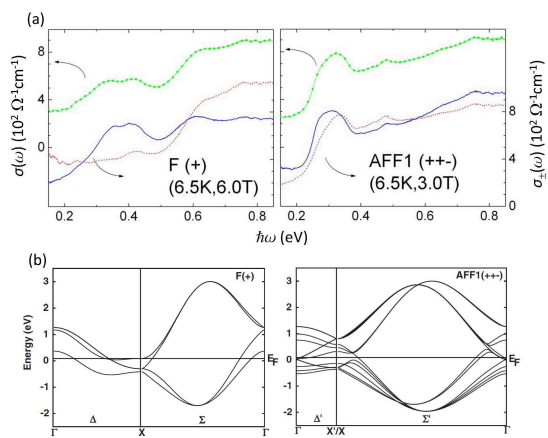}
\end{center}
\caption{
(Color online)
(a) Normal optical conductivity [$\sigma(\omega)$] using unpolarized light (solid circles) and circular-polarization-dependent optical conductivity [$\sigma_{\pm}(\omega)$] spectra (solid line for $+$, dashed line for $-$) of CeSb in the AFF1 and F phases. 
(b) Band structures of the F and AFF1 phases.
$X$ and $X'$ correspond to $(001)$ and $(001/3)$, respectively.
The $\Delta$ and $\Sigma$ axes are shown in the figure.
In the F phase, Ce~$5d$ bands have the highest energy at the $\Gamma$ point.
Ce~$5d$ bands are doubly degenerate in the $\Delta$ axis.
The other two bands on the axis are the Sb~$5p$ bands.
One of the Sb~$5p$ bands is pushed up by the strong $p$-$f$ mixing.
From Kimura {\it et al.}, 2002.~\cite{Kimura2002}
}
\label{CeSbIRMCD}
\end{figure}
The circular-polarization-dependent optical conductivity [$\sigma_{\pm}(\omega)$] spectra reflect the electronic polarization perpendicular to the magnetic field.
The $p$-$f$ mixing effect mainly works in the MCD configuration.
Figure~\ref{CeSbIRMCD}(a) shows the $\sigma(\omega)$ and $\sigma_{\pm}(\omega)$ spectra in the AFF1 and F phases.
The $\sigma_{\pm}(\omega)$ spectra were derived by the KK analysis of the obtained $R(\omega)$ spectra with circular polarized light.~\cite{DresselGruner}
The $\sigma_{\pm}(\omega)$ spectra in the F phase show a large MCD.
The $\sigma_{+}(\omega)$ spectrum has a peak at 0.35~eV compared with the $\sigma_{-}(\omega)$ spectrum, which has no obvious peaks and is similar to $\sigma(\omega)$ in the P phase.~\cite{Kimura2002}

The calculated band structures in the F and AFF1 phases including the $p$-$f$ mixing and $p$-$d$ mixing effects are shown in Fig.~\ref{CeSbIRMCD}(b).
This result suggests that the band structure is strongly affected by the mixing effect and also by the alignment of magnetic moments.
The peaks observed in $\sigma_{\pm}(\omega)$ spectra in the ordered state reflect the complex band structure.

\subsection{IR spectroscopy under multi-extreme conditions: pseudo gap formation and collapse in CeSb}

Using IR/THz-SR, spectroscopy under multi-extreme conditions of low temperature, high pressure, and high magnetic field can be performed owing to its high brilliance property.
The physical properties of SCES can be markedly changed by such external conditions.
To obtain information on the origin of the physical properties, an optical investigation of the electronic structure for the same sample condition must be performed.
Nishi {\it et al.} designed the first IR spectroscopy under multi-extreme conditions of the combination of low temperature, high pressure, and high magnetic field using IR-SR, and applied the method to the anomalous high-pressure phase of CeSb.~\cite{Nishi2005-1}
This work is reviewed below.

When an external pressure above about 2~GPa is applied to CeSb, the electrical resistivity ($\rho\sim$ several~m$\Omega\cdot$cm) at around $T=$~30~K increases by one full order over that at ambient pressure ($\rho\sim$ 140~$\mu\Omega\cdot$cm).~\cite{Mori1993}
The magnetic phase, in which the enhancement appears, is the single-layered antiferromagnetic (AF-1) phase with the magnetic structure $+-$, which is not present at ambient pressure.~\cite{Osakabe2003}
The difference between the AF-1A and AF-1 phases is in the magnetic structure, but $\rho$ in the AF-1A phase ($\rho\leqq$10~m$\Omega\cdot$cm) is that of a metallic phase, which is also different from that in the AF-1 phase.
The clarification of the difference between the electronic structures of the AF-1 and AF-1A phases and their temperature and magnetic field dependences is fundamental to understand the changes in the Sb~$5p$ band due to the difference in the magnetic structure.

\begin{figure}[b]
\begin{center}
\includegraphics[width=0.45\textwidth]{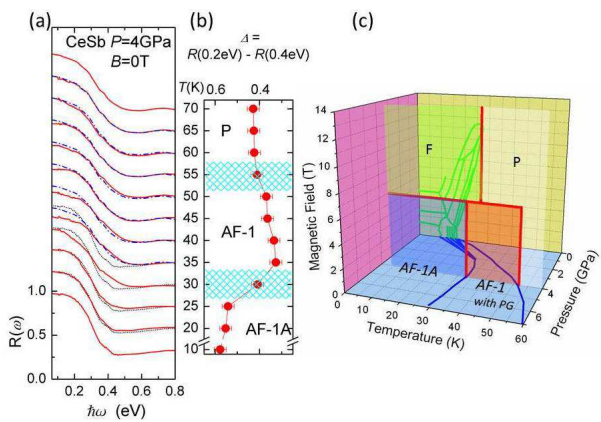}
\end{center}
\caption{
(Color online)
(a) Temperature dependence of the reflectivity [$R(\omega)$] spectrum of CeSb at $P=$~4~GPa and $B=$~0~T.
Successive curves are offset by 0.25 for clarity.
The spectra at 10 and 70~K are also plotted by dotted and dot-dashed lines, respectively, in the intermediate temperature range as a guide.
(b) The difference ($\Delta$) between the $R(\omega)$ intensity at 0.2~eV and that at 0.4~eV as a function of temperature.
The hatched areas are the phase transition temperatures.
The phase names of AF-1A, AF-1, and P follow the same notation as those for neutron scattering.
(c) Magnetic field--temperature ($B-T$) phase diagram of CeSb at $P=$~4~GPa derived from $\sigma(\omega)$ spectra under multi-extreme conditions.
The phase diagrams of $P-T$ at 0~T~\cite{Osakabe2003} and $B-T$ at ambient pressure~\cite{Kimura2002} are also plotted for the reference. AF-1 and AF-1A are the name of the phases.
Only AF-1 phase has a pseudo gap (PG) structure.
From Nishi {\it et al.}, 2005.~\cite{Nishi2005-1}
}
\label{CeSbphase}
\end{figure}
To perform IR spectroscopy under multi-extreme conditions, a diamond anvil pressure cell with a sample was set at the sample position located at the center of a superconducting magnet.
The sample conditions were $T\geq$~4~K, $P=$~4~GPa and magnetic fields of $B\leq$~14~T.
The temperature dependence of the $R(\omega)$ spectrum at $P=$~4~GPa and $B=$~0~T is shown in Fig.~\ref{CeSbphase}(a).
Below $T=$~30~K and above 60~K, the spectra are typically metallic because the $R(\omega)$ intensity approaches unity with decreasing photon energy.
At intermediate temperatures, the $R(\omega)$ spectrum displays a strong temperature dependence.
At 30~K, the spectrum changes markedly.
In particular, $R(\omega)$ becomes flat below 0.1~eV.
This means that the metallic character is suppressed at intermediate temperatures and a pseudo gap appears.

The pseudo gap collapses not only by increasing temperature but also by applying a magnetic field higher than 6.5$\pm$0.5 T.
The magnetic field - temperature ($B-T$) phase diagram at 4~GPa resulting from changes in the $R(\omega)$ spectrum is shown in Fig.~\ref{CeSbphase}(b).
In the phase (named as AF-1), an antiferromagnetic spin aligns directly along the magnetic field.
The Sb~$5p$ band is modulated by the magnetic structure and then the energy gap opens at $E_{\rm F}$.
In the figure, two phase diagrams of $P-T$ at $B=$~0~T~\cite{Osakabe2003} and $B-T$ at ambient pressure~\cite{Kimura2002} are also plotted.
The phase diagram at 4~GPa is simpler than that at ambient pressure.
In particular, the complex magnetic structure at ambient pressure disappears at 4~GPa.
At ambient pressure, since $p$-$f$ mixing competes with other magnetic interactions and crystal field splitting, such complex magnetic phases and structures appear.
With increasing pressure, $p$-$f$ mixing is enhanced and then becomes predominant among these interactions.
The enhancement of $p$-$f$ mixing also makes the simpler magnetic phase diagram at 4~GPa.
This is the plausible result of the $p$-$f$ mixing enhancement due to applied pressure.
This is the first optical observation of a magnetic-field-induced nonmetal-metal phase transition at high pressures.~\cite{Nishi2005-1}

\section{IR Microspectroscopy and Imaging under Extreme Conditions}\label{sec:imaging}

In conventional optical experiments, the measurement area of a sample is assumed to be uniform.  However, when the sample shows inhomogeneity, spatially resolved measurements, such as the mapping or imaging of the electronic structure, are important.  Sometimes, in SCES, the physical properties can be easily changed by external pressure and magnetic field because of such inhomogenity.  The high-brilliance property of IR/THz-SR can be used not only for the spectroscopy under extreme conditions but also for the spatial imaging of inhomogeneous materials.  It is also useful for very small samples because IR-SR can be focused in the diffraction limit. 

Recently, electronic inhomogeneities near Mott transition in transition metal oxides have been investigated by spectroscopic imaging such as photoemission scanning microscopy~\cite{lupi-2012,Sarma2004} and IR scanning near-field microscopy.\cite{Qazilbash2007}   Although the spatial resolution of IR microscopy is much lower than those of the above techniques, IR microscopy can be performed under high pressure and magnetic field, as already discussed.   In contrast, photoemission and IR near-field microscopies are technically difficult to perform under high pressure and magnetic field.   This advantage of IR microscopy is well realized when it is used to study electronic inhomogeneities in organic compounds, because organic compounds show electronic inhomogeneities with larger length scales than the oxides, as discussed later.  

Here, the IR microspectroscopy of minute samples such as graphene and carbon nanotube, and IR imaging of the phase separation in organic conductors and europium monoxide under a magnetic field and pressure are reviewed.  

\subsection{Graphene, carbon nanotubes, and organic FETs}

SR-based IR microspectroscopy has also been applied in the carbon-based and organic-conductor-based nanosciences.
For example, monolayer and bilayer graphenes,~\cite{heinz-2008a,basov-2008a,basov-2008b,basov-2009,martin-2009,heinz-2010a} single walled carbon nanotubes (SWCNTs),~\cite{heinz-2010b} and organic field effect transistors (FETs)\cite{martin-2006,martin-2007a,martin-2007b} have been studied.  
The IR spectra of these systems have revealed much information about, for example, their fundamental electronic structures and band gap, the dynamics and density of mobile carriers, and phonon and molecular vibrations.
A small sample of graphene was either mounted alone on a substrate,~\cite{heinz-2008a,heinz-2010a} or attached to metal electrodes and/or gates that were lithographically patterned on a substrate.~\cite{basov-2008a,basov-2008b,basov-2009,martin-2009}  The SWCNT work was also performed with electrodes attached to the sample.\cite{heinz-2010b}   The electronic excitations and the carrier dynamics in very narrow conducting channels of organic FETs were studied in detail.\cite{martin-2006,martin-2007a,martin-2007b}   With the small dimensions of these samples, the advantage of high-brilliance IR-SR is apparent.  Below, some of the works on graphene and SWCNTs are briefly reviewed.  

Mak {\it et al.} reported the $\sigma(\omega)$ of monolayer graphene obtained from measured reflectance and transmittance in the 0.2--1.2~eV range.~\cite{heinz-2008a} They used a graphene sample obtained by the mechanical exfoliation of kish graphite and mounted it on a quartz substrate.  Then they identified a monolayer region, with a typical area of several hundreds to several thousands of $\mu$m$^2$, and studied it by IR-SR microspectroscopy.  They found that, in the 0.5--1.2~eV range, $\sigma(\omega)$ was constant within $\pm$10\% of the theoretically predicted universal value of $\pi G_0/4$, where $G_0 =2e^2/h$ is the quantum of conductance.  This remarkable result is a manifestation of the unique electronic structures of graphene with ``massless Dirac electrons.''  The conductance corresponds to a relative absorbance of 2.3\%, which appears remarkably large for the very small thickness of monolayer graphene.   Below 0.5~eV, however, $\sigma(\omega)$ notably deviated from its universal value, which was attributed to intraband electronic excitations.~\cite{heinz-2008a}

\begin{figure}[b]
\begin{center}
\includegraphics[width=0.4\textwidth]{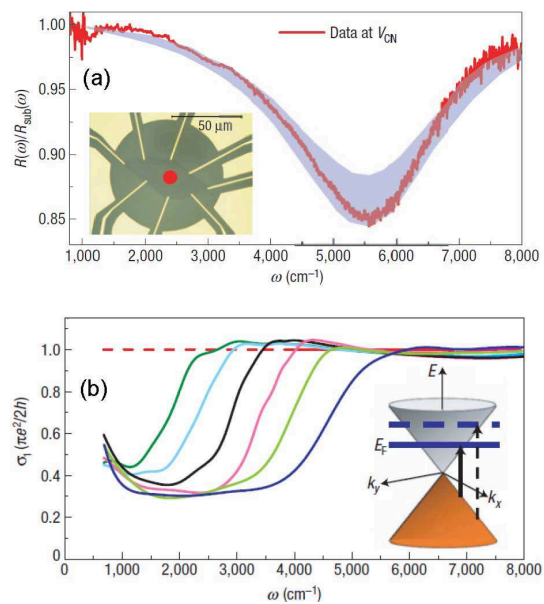}
\end{center}
\caption{
(Color online) 
Microspectroscopic study of a monolayer graphene by IR-SR reported by Li {\it et al.}~\cite{basov-2008a} (a) Reflectance [$R(\omega)$] spectrum of monolayer graphene device with lithographically patterned gate electrodes, pictured in the inset, divided by that of the substrate.  
The shaded area shows the range of reflectance expected from $\pm$15\% of the universal conductance $\pi e^2/2h$.
In the inset, the red spot at the center of the sample indicates the size of the focused IR-SR beam.
(b) Optical conductivities ($\sigma_1$) of graphene at different gate voltages relative to the charge neutrality voltage, which places the Fermi level ($E_{\rm F}$) at the crossing point of the Dirac cone ($E$=0 in the inset).
The applied voltage changes from 10~V (left) to 71~V (right).
Reprinted with permission from Macmillan Publishers Ltd: 
Nature Physics {\bf 4} (2008) 532, copyright 2008.  
}
\label{fig:graphene}
\end{figure}
Monolayer and bilayer graphenes under an electric field have been studied by IR-SR-based microspectroscopy.~\cite{basov-2008a,basov-2008b,basov-2009,martin-2009}  These works used a graphene sample mounted with a lithographically patterned gate electrode, and the electric field was applied with a gate voltage.   Figure~\ref{fig:graphene} shows an example of an experiment on monolayer graphene.~\cite{basov-2008a}  By tuning the gate voltage, the carrier density was controlled, and hence the Fermi level ($E_{\rm F}$) was tuned across the crossing point of Dirac cone, depicted in the inset of Fig.~\ref{fig:graphene}.  (A charge neutral state is realized when $E_{\rm F}$ is located at the crossing point.)  The measured $\sigma(\omega)$ shows a universal conductance behavior, which was already mentioned above, at photon energies corresponding to interband transitions to those above $E_{\rm F}$ [the broken vertical arrow in the inset of Fig.~\ref{fig:graphene}(b)].  However, at lower energies where interband excitations are blocked by occupied states, $\sigma(\omega)$ is seen to deviate from the universal value due to intraband electronic excitations.  The shift of the $\sigma_1$ [$=\sigma(\omega)$] curve to a higher frequency with the gate voltage in Fig.~\ref{fig:graphene}(b) indicates the tuning of $E_{\rm F}$ with the gate voltage.  Bilayer graphenes were also studied using similar devices and IR-SR.\cite{basov-2008b,basov-2009,martin-2009}  For example, it was shown that the magnitude of the band gap in bilayer graphene was widely tunable by changing the gate voltage, showing potential for future applications in opto-electronics.\cite{martin-2009}

\begin{figure}[b]
\begin{center}
\includegraphics[width=0.45\textwidth]{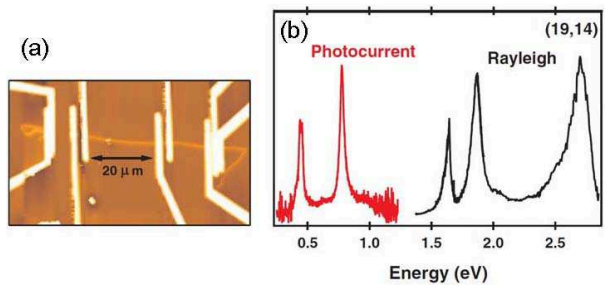}
\end{center}
\caption{
(Color online) 
Photocurrent (PC) study of an individual, large-diameter, semiconducting single-walled carbon nanotube (SWCNT) reported by Sfeir {\it et al.}~\cite{heinz-2010b}   
(a) Image of a SWCNT sample attached to lithographically defined source and drain electrodes.  
The diameter of the SWCNT, seen horizontally in the image, was a few nm.   
The SWCNT was illuminated by the focused spot of IR-SR, and the resulting PC was recorded by an FTIR interferometer to derive the PC spectrum. 
(b) Measured PC spectrum for a SWCNT having a diameter of 2.3~nm, together with the Rayleigh scattering spectrum.   
Five lowest subband transitions are observed (two from PC, three from Rayleigh).  (19,14) denotes the chirality of the SWCNT sample.  
Figure reprinted with permission from Sfeir {\it et al.}, Phys. Rev. B {\bf 82}, 195424, 2010.
Copyright 2010 by the American Physical Society.  
}
\label{fig:CNT}
\end{figure}
SWCNT has also been studied by IR-SR-based microspectroscopy.  
Sfeir {\it et al}. used a semiconducting, large-diameter SWCNT device attached to lithographically patterned source and drain electrodes, as shown in Fig.~\ref{fig:CNT},~\cite{heinz-2010b} to study electronic transitions.  The diameter of the SWCNT was only a few nm, with the distance between electrodes about 20~$\mu$m.  The effective area of the sample facing the incoming IR light is so small that the study would have been difficult without the high brilliance of IR-SR.  However, the diameter of SWCNT was apparently too small for a reflectance/transmittance study even with the IR-SR.  Instead, the photocurrent (PC) across the SWCNT was measured with the electrodes as a function of the mirror displacement in the FTIR.  The resulting interferogram was Fourier-transformed to obtain PC spectra.  Figure~\ref{fig:CNT}(b) shows a PC spectrum of a SWCNT with a 2.3~nm diameter, together with the Rayleigh scattering spectrum of the same sample.  The peak at 0.48~eV is due to the excitonic transition between the two lowest electron and hole subbands.   The subbands up to fifth-lowest energy (two from PC, three from Rayleigh) are clearly identified.

\subsection{IR imaging of organic superconductors under extreme conditions}


\begin{figure}[b]
\begin{center}
\includegraphics[width=0.45\textwidth]{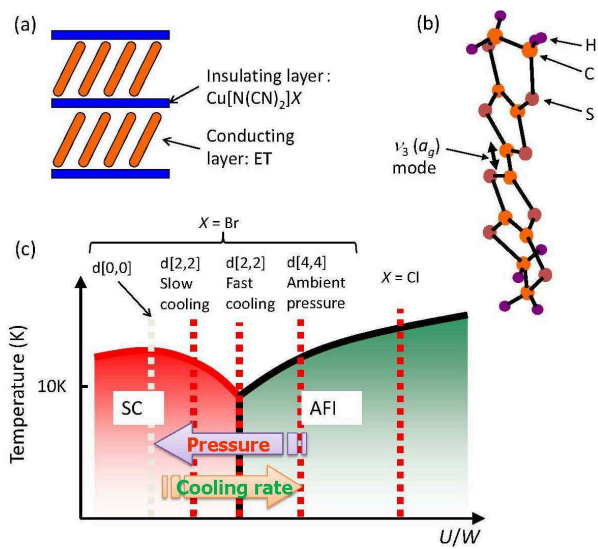}
\end{center}
\caption{
(Color online)
(a) Schematic figure of crystal structure of quasi-two-dimensional organic conductors $\kappa$-(ET)$_2$Cu[N(CN)$_2$]$X$ ($X=$~Br, Cl).
(b) Structure of ET molecule.
(c) Schematic phase diagram of $\kappa$-(ET)$_2$Cu[N(CN)$_2$]$X$ ($X=$~Br, Cl) as a function of the ratio of the on-site Coulomb interaction energy $U$ to the band width $W$.
$X=$~Br is located close to the Mott transition boundary; on the other hand, $X=$~Cl is located in the antiferromagnetic insulator phase far from the boundary.
Applying pressure and increasing cooling rate are consistent with the increase and decrease in $U/W$, respectively.
}
\label{ETphase}
\end{figure}
The quasi-two-dimensional organic conductor $\kappa$-(ET)$_2$Cu[N(CN)$_2$]$X$ ($X=$Br, Cl), in which ET molecule is called bis(ethylenedithio)-tetrathiafulvalene (BEDT-TTF, Fig.~\ref{ETphase}b), has a layered crystal structure shown in Fig.~\ref{ETphase}(a).~\cite{Ishiguro1998}
The conducting ET layer and insulating Cu[N(CN)$_2$]$X$ layer interlaminate.
In these materials, the physical properties can be controlled by the change of the insulating layer.
For instance, $X=$~Cl (denoted $\kappa$-Cl hereafter) is an antiferromagnetic insulator (AFI) but $X=$~Br (denoted $\kappa$-Br hereafter) changes to a superconductor (SC) in the ground state.
In the $\kappa$-Br case, the change of the side chain of the ET molecule from hydrogen to deuterium~\cite{Kawamoto1997} or the cooling rate~\cite{Taniguchi1999} induces the SC-to-AFI transition.
The fast cooling induces the disorder of ethylene end groups.~\cite{Geiser1991}
It is important that such small perturbations induce a marked change from SC to AFI in the electronic state.
The origin is the relationship between the bandwidth $W$ (the kinetic energy of conduction electrons) and the on-site Coulomb interaction $U$, that is, the Mott transition.
$\kappa$-(ET)$_2$Cu[N(CN)$_2$]$X$ can be well explained by the phase diagram shown in Fig.~\ref{ETphase}(c).~\cite{Kanoda2006}
By using a partial substitution of hydrogen of the side chain of the ET molecule to deuterium, the ground state in the vicinity of the Mott transition boundary can be controlled in detail.
For instance, for 50\% substitution, namely, $d[2,2]$, the ground state is SC when the sample is cooled down slowly; on the other hand, the ground state shifts to the borderline of the Mott transition when the sample is cooled down rapidly.
For 75\% substitution ($d[3,3]$), the slow-cooling case occurs just at the Mott transition boundary, but it shifts to AFI in the case of fast cooling.~\cite{Kawamoto1998}
Since the ground state as well as the physical properties can be changed by the small perturbation, these materials are suitable for the investigation of the mechanism of the Mott transition and the change in physical properties.

\begin{figure}[b]
\begin{center}
\includegraphics[width=0.45\textwidth]{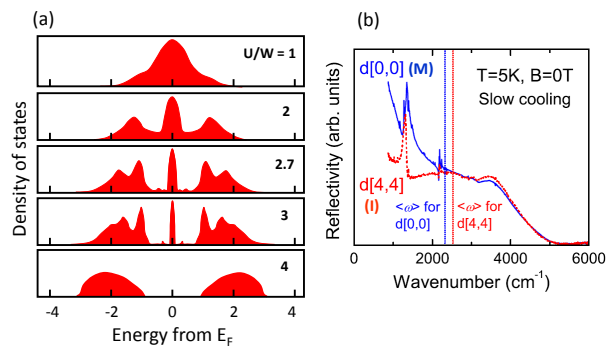}
\end{center}
\caption{
(Color online)
(a) Density of states as a function of the ratio of the on-site Coulomb interaction energy $U$ to the bandwidth $W$ calculated using the dynamical mean-field theory.~\cite{Zhang1993}
(b) Typical reflectivity [$R(\omega)$] spectra in the superconducting ($d[0,0]$, $T$~=~5~K, $B$~=~0~T) and antiferromagnetic insulating ($d[4,4]$, $T$~=~5~K, $B$~=~0~T) phases and their centers of spectral weights ($\langle\omega\rangle$).
}
\label{MottDOS}
\end{figure}
The density of states (DOS) that explains the Mott transition is shown in Fig.~\ref{MottDOS}(a), which was calculated by a quantum Monte-Carlo method.~\cite{Zhang1993}
The figure indicates that DOS in metallic states accumulates near $E_{\rm F}$; on the other hand, the main part of DOS moves far away from $E_{\rm F}$.
When the change in electronic structure is realized, the center of gravity ($\langle\omega\rangle$) of the $R(\omega)$ spectrum that corresponds to joint DOS shifts from the low-energy side in the metallic state to the high-energy side in the insulating state (Fig.~\ref{MottDOS}b).
Then the change in $\langle\omega\rangle$ indicates the metal--insulator transition.

To acquire spatial imaging data, a total of 931 spectra were acquired in a 360$\times$360~$\mu$m$^{2}$ region in steps of 12~$\mu$m with a spatial resolution of less than 20~$\mu$m.
All spectra were recorded with unpolarized light because the overall spectral change corresponding to each phase is focused.
The spatial image of the $R(\omega)$ spectra was plotted using the $\langle\omega\rangle$ obtained using the following function:~\cite{Nishi2005-2}
\begin{equation}
\langle\omega\rangle = \int_{\omega_1}^{\omega_2} \omega R(\omega) d\omega/ \int_{\omega_1}^{\omega_2} R(\omega) d\omega ,
\label{eq:COG}
\end{equation}
where $\omega_1$ is the lowest accessible wavenumber, 870~cm$^{-1}$, and $\omega_{2}$ is set to 5000~cm$^{-1}$, above which there is no difference between the metallic and nonmetallic spectra~\cite{Gri1999}.
The reason for this convergence in the spectra is that the change in the electronic structure due to the Mott transition~\cite{Zhang1993} results in a shift in the optical spectral weight, because the optical spectra indicate the relative energy difference between the valence and conduction bands.
For instance, typical $R(\omega)$ spectra in the SC ($d[0,0]$, $T$~=~5~K, $B$~=~0~T) and AFI ($d[4,4]$, $T$~=~5~K, $B$~=~0~T) phases and their respective $\langle\omega\rangle$ values are shown in Fig.~\ref{MottDOS}(b).
The difference in the electronic structure is directly reflected in the shape of $R(\omega)$ spectra.
Therefore, $\langle\omega\rangle$ was employed as a representative of the shape of $R(\omega)$ as well as the electronic structure.
Note that the $R(\omega)$ spectrum in the SC phase is different from a normal metallic one only in the THz region of the energy region with less than two or four times higher energy than the SC transition temperature.
It is impossible to distinguish between the SC and normal metallic spectra in the infrared region.
Then, the observed metallic state is regarded as the SC state below the critical magnetic field $H_{c2}$ ($\sim$~4~T) hereafter.
Note that a similar metal-insulator phase separation can be detected using the energy shift of the $\nu_3 (e_g)$ molecular vibration mode in the ET molecule shown in Fig.~\ref{ETphase}(b).~\cite{Sasaki2004}

\subsubsection{Magnetic-field-induced Mott transition}

In the presence of magnetic fields, the 50\%-deuterated $\kappa$-Br ($d[2,2]$) under the fast-cooling conditions directly changes from the SC state to the AFI state in spite of the fact that the SC state of the material formed under slow-cooling conditions normally changes to a paramagnetic metallic (PM) state with increasing magnetic-field strength.~\cite{Taniguchi2003}
The change has been investigated by macroscopic electrical resistivity measurement, but the origin has not been determined yet.
Then IR imaging under a magnetic field was performed to investigate the origin of the SC-to-AFI transition induced by a magnetic field as well as by the phase separation.

\begin{figure}[b]
\begin{center}
\includegraphics[width=0.45\textwidth]{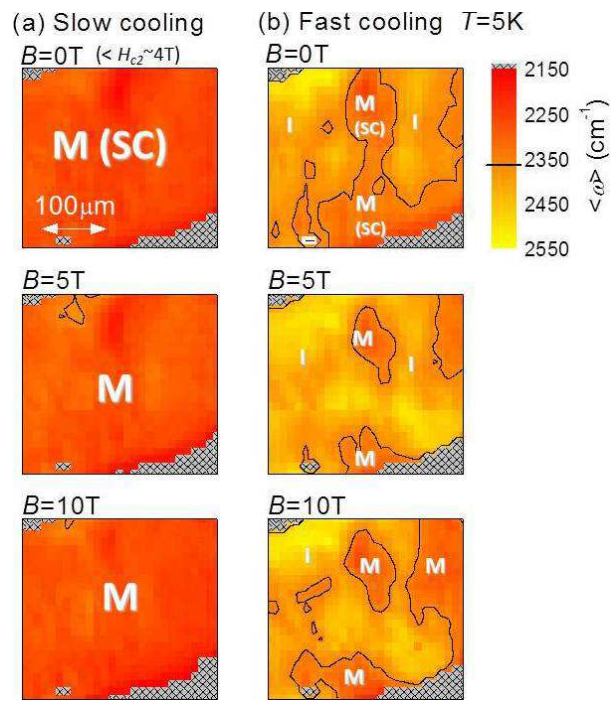}
\end{center}
\caption{
(Color online)
Magnetic field strength and cooling rate [slow cooling in (a) and fast cooling in (b)] dependences of the spatial imaging of the center of spectral weight ($\langle\omega\rangle$) of 50\%-deuterated $\kappa$-Br ($d[2,2]$) at $T$~=~5~K.
The black lines indicate the rough M--I boundary ($\omega_{MI}$) of 2,350~cm$^{-1}$, and the lower and higher wavenumbers indicate the insulating and metallic (superconducting) reflectivity spectra.
The hatched area is the area outside of the sample.
From Nishi {\it et al.}, 2007.~\cite{Nishi2007}
}
\label{d22image}
\end{figure}
Figure~\ref{d22image} shows the magnetic field and cooling rate dependences of the spatial distribution of $\langle\omega\rangle$ of the metallic and insulating states.
To observe the cooling rate dependence of the samples, two cooling rates, i.e., $-17$~K/min for fast cooling, and $-0.05$~K/min for slow cooling, were used in the temperature range from 90--70~K.
The IR $R(\omega)$ imaging was performed in the wavenumber range from 870 to 8000~cm$^{-1}$ at 5~K at magnetic fields of 0, 5, and 10~T.

In the case of slow cooling at 0~T, even though the spatial image is not monochromatic, the $\langle\omega\rangle$ over the entire sample surface is lower than that at the rough boundary ($\omega_{MI}=$~2350~cm$^{-1}$), which was evaluated elsewhere.~\cite{Nishi2005-2}
This indicates that the whole sample surface is in the metallic state, which is consistent with the electrical resistivity and ac-susceptibility data.~\cite{Taniguchi1999}
The metallic state does not change with increasing magnetic field strength of up to 10~T.
This result is also consistent with the electrical resistivity data under magnetic fields.~\cite{Taniguchi2003}

In the case of the fast-cooling conditions at 0~T, $\langle\omega\rangle$ shifts to the higher wavenumber side.
This indicates that the insulating region expands compared with the results obtained under the slow-cooling conditions.
With increasing magnetic field strength up to 5~T, $\langle\omega\rangle$ shifts to the higher wavenumber side and the insulating region expands in contrast to the constant $\langle\omega\rangle$ distribution observed in the slow-cooling experiment.
This result is consistent with the electrical resistivity data if the following explanation is accurate.
At 0~T, SC domains connect to one another resulting in the electrical resistivity dropping to zero even if AFI domains exist.
Actually, the percolation of the SC state appears in the top figure of Fig.~\ref{d22image}(b).
The coexistence of SC and AFI states is also consistent with the ac-susceptibility data that does not show perfect diamagnetism.~\cite{Taniguchi1999}
When magnetic fields are applied, AFI domains expand and SC (or metallic) domains disconnect as shown in the middle panel of Fig.~\ref{d22image}(b).
As a result, the electrical resistivity markedly changes to reflect an insulating state.
Actually, the reentrant SC phase appears at the boundary of the AFI--SC transition in $\kappa$-Cl~\cite{Ito1996} and $d[4,4]$~\cite{Ito2000} with increasing pressure.
The origin of this observation has been revealed to be a mixture of SC and AFI domains resulting from the pressure-dependent electrical resistivity of $\kappa$-Cl~\cite{Kagawa2004, Kagawa-PRL2004} and from the pressure-dependent $R(\omega)$ of $d[4,4]$ in the infrared region~\cite{Kimura2007}.
Therefore, the SC--AFI transition observed in the electrical resistivity measurement originates from the increase in the size of the AFI region in the sample.
The coexistence of SC and AFI domains has been studied using the half-filled Hubbard model in the limit of infinite dimensions~\cite{Laloux1994}.
In this paper, we pointed out that the magnetic field-induced M--I transition occurs in the vicinity of $U/W\sim1$.
$d[2,2]$ is only an example.
Therefore, the origin of the SC--AFI transition in $d[2,2]$ is concluded to be the predicted M--I transition in which the critical magnetic field of the Mott transition is below $H_{c2}$.~\cite{Nishi2007}

\subsubsection{Pressure-induced Mott transition}

The ground state of perfectly deuterated $\kappa$-Br ($d[4,4]$) is located in the AFI state, as shown in Fig.~\ref{ETphase}.
When pressure is applied to the material, the bandwidth $W$ increases and the ground state changes to SC.
For instance, the pressure-dependent electrical resistivity demonstrates that the insulating behavior appears at low pressures; however, the electrical resistivity drops to zero at low temperatures when a pressure of about 3~MPa is applied.~\cite{Ito2000} 
However, a reentrant electrical resistivity appears, in which the electrical resistivity changes to insulating at temperatures lower than the SC transition temperature.
To clarify the origin of the reentrant electrical resistivity, infrared $R(\omega)$ imaging was performed under pressure.~\cite{Kimura2007}

\begin{figure}[b]
\begin{center}
\includegraphics[width=0.45\textwidth]{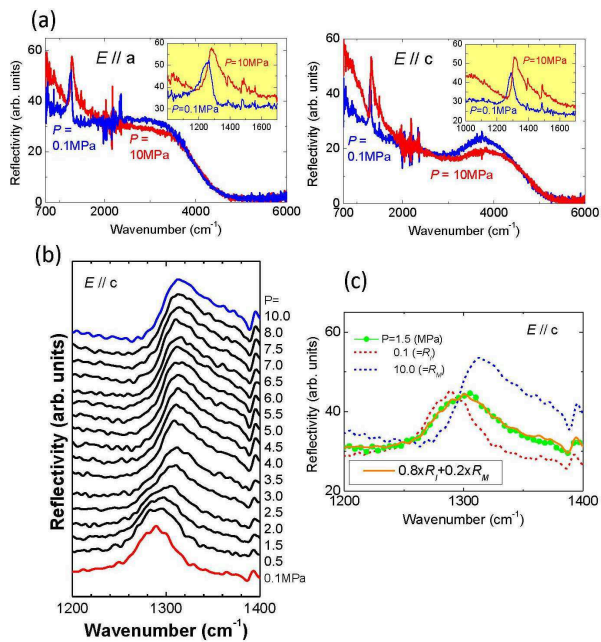}
\end{center}
\caption{
(Color online)
(a) Pressure-dependent polarized $R(\omega)$ spectra of perfectly deuterated $\kappa$-Br ($d[4,4]$) at 5~K. 
The insets are the $\nu_3 (e_g)$ vibration modes on a larger scale.
(b) Pressure-dependent spectral change in the $\nu_3 (e_g)$ vibration mode with the polarization of $E \| c$.
(c) The fitting result of the spectrum at $P=$~1.5~MPa with a linear convolution of $R(\omega, {\rm 1.5~GPa})=0.8 \times R(\omega, {\rm 0.1~GPa})+0.2 \times R(\omega, {\rm 10~GPa})$.
From Kimura {\it et al.}, 2007.~\cite{Kimura2007}
}
\label{d44}
\end{figure}
In ordinal infrared spectroscopy under pressure, DAC described before is generally used.
However, the available pressure of DAC is 0.1 to several 10~GPa, the minimum step of the pressure is larger than several tens of MPa.
In the present experiment, since a maximum of 10~MPa is needed, DAC cannot be used.
Then a new pressure cell was developed, in which the pressure is controlled by adjusting the pressure directly from a helium cylinder.~\cite{Nishi2006}

The pressure-dependent polarized $R(\omega)$ spectra of $d[4,4]$ at pressures of 0.1 and 10~MPa are shown in Fig.~\ref{d44}(a).
In the experiment, the spatial resolution is about 50~$\mu$m, that is, the spectra are the average in the region.
At 0.1~MPa, the polarized $R(\omega)$ spectra in both polarizations of $E \| c$ and $E \perp c$ are insulating because no Drude-like increase toward zero frequency appears.
At 10~MPa, on the other hand, the intensities at 3500 and below 2000~cm$^{-1}$ decreases and increases, respectively, indicating a metallic spectrum.

The inset of Fig.~\ref{d44}(a) is the peak of the $\nu_3 (e_g)$ vibration mode in the ET molecule at 1290~cm$^{-1}$ (Fig.~\ref{ETphase}b).
At ambient pressure, the peak wavenumber is 1290~cm$^{-1}$, whose peak shape and wavenumber are the same as those previously reported~\cite{Gri1999}.
By applying pressure from an insulating phase to a metallic phase, the peak changes from a symmetric peak to a Fano-like asymmetric peak that has a tail to the higher-wavenumber side, owing to the interaction between the $\nu_3 (e_g)$ vibration mode and associated carriers.
The spectral shape and peak wavenumber are the same as those of non-deuterated $\kappa$-Br in the SC phase.
This is a direct optical observation of the pressure-induced AFI-to-SC transition induced by pressure.
At intermediate pressures, the spectra can be explained by linear combinations of the spectra at 0.1 and 10~MPa.
This implies the coexistence of SC and AFI in $d[4,4]$ at intermediate pressures.
This behavior is the same as that in $\kappa$-Cl at around 30~MPa~\cite{Lefebvre2000,Kagawa-PRL2004,Kagawa2004}.
Additionally, the peak shifts to the higher-wavenumber side with phase changes to the metallic one.
The peak shape and wavenumber indicate the pressure-dependent insulator-to-metal transition.
For instance, the pressure-dependent peak of the $\nu_3 (e_g)$ vibration mode at 5~K is shown in Fig.~\ref{d44}(b).
The peak wavenumber shifts to the higher-wavenumber side and the shape changes to an asymmetric one.
The crossover pressure is about 2~MPa, i.e., the pressure is the boundary of AFI and SC.
However, the spectrum changes continuously.
This is inconsistent with the Mott transition, which is considered to be a first-order transition with a discontinuous change in electronic structure.

\begin{figure}[b]
\begin{center}
\includegraphics[width=0.45\textwidth]{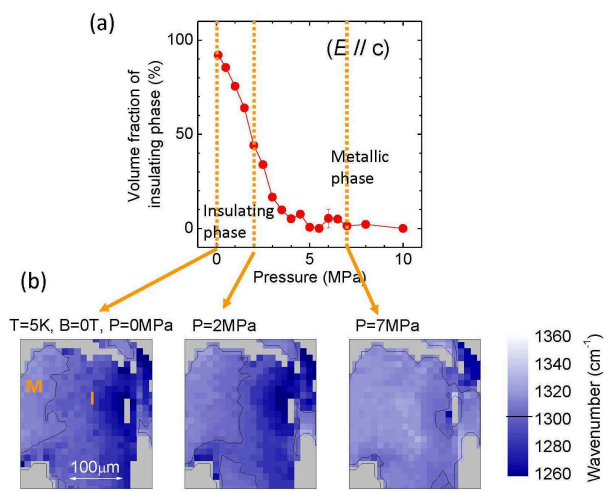}
\end{center}
\caption{
(Color online)
(a) Pressure dependence of the volume fraction of the AFI phase $p$-$f$ $d[4,4]$ obtained by the fitting of the $\nu_3 (e_g)$ modes in Fig.~\ref{d44}.
(b) Pressure dependence of the spatial imaging of the center of spectral weight ($\langle\omega\rangle$) of perfectly deuterated $\kappa$-Br ($d[4,4]$) at $T$~=~5~K.
The black lines indicate the rough M--I boundary ($\omega_{MI}$) of 2,350~cm$^{-1}$, and the lower and higher wavenumbers indicate the insulating and metallic (superconducting) reflectivity spectra.
The hatched area is the area outside of the sample.
From Kimura {\it et al.}, 2008.~\cite{Kimura2008b}
}
\label{d44image}
\end{figure}
The volume fraction of the AFI phase can be evaluated using linear combinations of the spectra at 0.1 and 10~MPa, as shown in the inset of Fig.~\ref{d44}(c).
After applying the same evaluation for all measurements, the volume fraction of the AFI phase was plotted as a function of pressure, as shown in Fig.~\ref{d44image}(a).
The figure indicates that the volume fraction of the AFI phase gradually decreases with increasing pressure, reaching almost 0\% above 5~MPa.
The transition pressure region is qualitatively consistent with the reentrant superconducting phase previously reported.~\cite{Ito2000}
On the basis of these results, the reentrant superconducting phase is concluded to originate from the phase coexistence of AFI and SC.~\cite{Kimura2007}

To determine the size of phase separation, the pressure-dependent spatial imaging of the $d[4,4]$ sample was performed.~\cite{Kimura2008b}
In detail, the spatial images at 0.1~MPa (mostly insulating), 7~MPa (mostly metallic), and 2~MPa (the intermediate pressure) were measured in 12~$\mu$m steps, and the peak wavenumber of the $\nu_3 (e_g)$ vibration mode was plotted in Fig.~\ref{d44}(e).
The borderline between the insulator and the metal was set at the center wavenumber (1300~cm$^{-1}$) of these peaks at 0.1 and 10~MPa.
At 0~MPa, small area on the right side having a peak above 1300~cm$^{-1}$ is the metallic phase, but the major area is the insulating phase because peaks appear below 1300~cm$^{-1}$.
This result clearly indicates that the sample is inhomogeneous.
While applying pressure up to 2~MPa, the metallic area expands to about 50\%.
At 7~MPa, almost all the sample changes to metallic, which is consistent with the volume fraction described before.
The transition from the insulating phase to the metallic phase changed continuously despite the notion that the phase transition is discontinuous.
This suggests that the insulating and metallic domain sizes of the phase separation are much less than the spatial resolution of 12~$\mu$m.
Therefore a higher spatial resolution beyond the diffraction limit using a near-field microspectroscopy is needed to detect a clear phase separation.

\subsection{IR magneto-optical imaging on EuO}

In first-order transitions such as the magnetic transition in magnetic materials and the Mott transition in SCES, physical properties markedly change.
Around the transition point, the phase separation, in which both phases coexist, sometimes appears.
Anomalous physical properties often originate from the phase separation, for instance, the colossal magnetoresistance of manganese oxides~\cite{Zhang2002} and the magnetic-field-induced superconductor-insulator transition of organic superconductors~\cite{Nishi2007} are considered to originate from the phase separation.
The domain size of the phase separation is typically on the nanometer scale.
Owing to the lattice distortion or trapped magnetic polarons, however, micrometer-sized domains sometimes appear.
Such large domains and the electronic structure of each domain can be identified by IR imaging.
The target material dealt with in this study, namely, electron-doped europium monoxide (EuO), has a large magnetoresistance,~\cite{Mauger1986} whose origin is still debated to be trapped magnetic polarons.
Therefore, the detection of a large domain size as well as of the inhomogeneity of the sample surface near the ordering temperature/magnetic field is important for studying the magnetic polaron scenario.

EuO is a ferromagnetic semiconductor with a Curie temperature ($T_{\rm C}$) of around 70~K.~\cite{Miyazaki2009}
With excess Eu electron doping or the substitution of Gd$^{3+}$ or La$^{3+}$ for Eu$^{2+}$-ions, $T_{\rm C}$ increases up to 200~K~\cite{Miyazaki2010} and the electrical resistivity drops twelve orders of magnitude below $T_{\rm C}$.~\cite{Schafer1968}
Since the magnetic moment originates from local Eu$^{2+}$ $4f^7$ electrons, electron-doped EuO has larger magnetic moments than colossal magneto-resistance manganites, which exhibit a similar insulator-to-metal transition at $T_{\rm C}$.
Therefore, electron-doped EuO has been attracting attention as a next-generation functional material for spintronic devices.~\cite{Schmehl2011}

\begin{figure}[b]
\begin{center}
\includegraphics[width=0.40\textwidth]{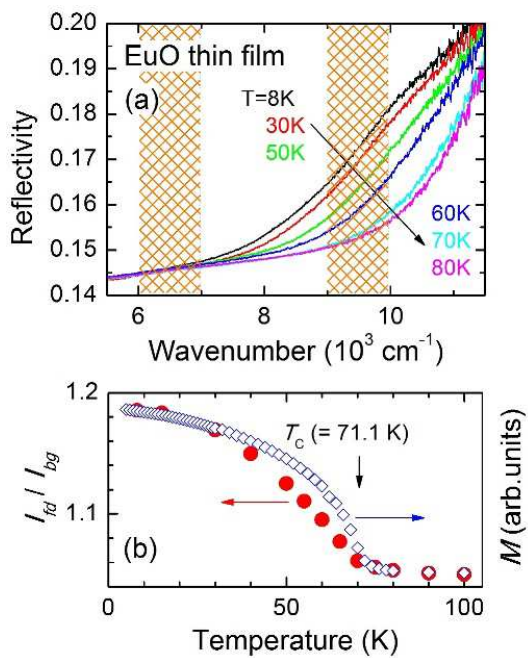}
\end{center}
\caption{
(Color online)
(a) Temperature dependence of the reflectivity [$R(\omega)$] spectrum of a EuO thin film in the wavenumber range of 4000--12000~cm$^{-1}$.
(b) The temperature dependence of the intensity ratio between the absorption edge of the exciton of the Eu~$4f \rightarrow 5d$ transition ($I_{fd}$) integrated at 9000--10000~cm$^{-1}$ and the background intensity ($I_{bg}$) at 6000--7000~cm$^{-1}$.
The temperature dependence of the magnetization (open diamonds) is also plotted.
From Kimura {\it et al.}, 2008.~\cite{Kimura2008a}
}
\label{EuOspectrum}
\end{figure}
To obtain information on electronic phase separation, IR $R(\omega)$ imaging using IR-SR with a diffraction limit spatial resolution was performed in the 6000 to 12000 cm$^{-1}$ wavenumber range at a 10~cm$^{-1}$ resolution at different temperatures from 40 to 80~K at magnetic fields of up to 5~T.
To acquire spatial imaging data, a total of 1681 spectra were obtained in a 200$\times$200~$\mu$m$^2$ region in steps of 5~$\mu$m with a spatial resolution higher than 5~$\mu$m.
The spatial images and the temperature dependence of the $R(\omega)$ spectrum were plotted using the intensity ratio of the absorption edge of the exciton of the Eu $4f \rightarrow 5d$ transition ($I_{fd}$) integrated over 9000--10000~cm$^{-1}$ to the background intensity ($I_{bg}$) in the 6000--7000~cm$^{-1}$ range, as shown in Fig.~\ref{EuOspectrum}(a).
The temperature-dependent energy gap shift is consistent with the magnetization due to the ferromagnetic transition, as shown in Fig.~\ref{EuOspectrum}(b).

\begin{figure}[b]
\begin{center}
\includegraphics[width=0.45\textwidth]{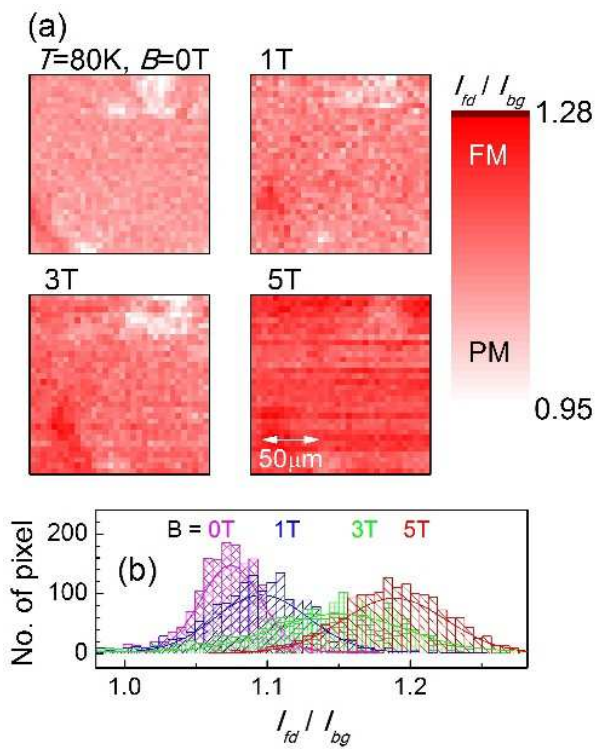}
\end{center}
\caption{
(Color online)
(a) Magnetic field dependence of $I_{fd}/I_{bg}$ of a EuO thin film at $T=$~80~K, which is slightly higher than $T_{\rm C}$.
(b) Statistical distributions and Gaussian fittings of $I_{fd}/I_{bg}$ derived from (a).
From Kimura {\it et al.}, 2008.~\cite{Kimura2008a}
}
\label{EuOimaging}
\end{figure}
To investigate the phase separation in the magnetic-field-induced paramagnetic-to-ferromagnetic transition, we performed infrared imaging at magnetic fields at 80~K, which is slightly higher temperature than $T_{\rm C}$, as shown in Fig.~\ref{EuOimaging}(a).
The figure indicates that the paramagnetic state at $B=$~0~T becomes ferromagnetic as the magnetic field increases.
A similar magnetic-field-induced insulator-to-metal transition has been observed in manganites in spite of its different origin.
The spatial distribution is plotted in Fig.~\ref{EuOimaging}(b).
The figure shows that the distribution width normalized by that at $B=$~0~T exhibits an approximately twofold increase at 3~T and then decreases at 5~T.
Such an anomalous temperature-dependent width can be explained in terms of the magnetic polaron scenario, in which a large magnetic polaron is generated by applying a magnetic field.~\cite{Yu2005,Yu2006}
In other words, the phase transition from a uniform paramagnetic state to a uniform ferromagnetic state via the inhomogeneous magnetic polaron state appears with increasing magnetic field at a temperature slightly higher than $T_{\rm C}$. 

These results are used as basis for discussing the domain size of the magnetic polaron state in EuO.
The spatial resolution of about 5~$\mu$m used in this work is larger than this domain size.
However, changes in the spatial distribution width can be detected.
In such cases, the domain size is about one-tenth of the spatial resolution of the microscope used, i.e., about several 100~nm.
This is the first direct observation of the inhomogeneity due to magnetic polaron domains.~\cite{Kimura2008a}

%
\section{Future IR/THz Light Sources Using Accelerator}\label{sec:CSR}

Up to now, the present status of IR/THz spectroscopy using normal SR has been introduced.
Using the high brilliance and polarization properties, some advanced spectroscopies under extreme conditions become available.
One of the areas of accelerator-based IR spectroscopy still being developed is the use of IR/THz free-electron lasers (FELs) and CSR.

IR/THz-FELs are widely used for electron spin resonance (ESR) and other methods.
In Japan, IR FELs have been installed at Tokyo University of Science~\cite{Imai2010} and Kyoto University~\cite{Zen2008} and THz-FEL at Osaka University~\cite{Isoyama2008}.  
Main aims of these FELs are the development of light sources, and there have been only a few examples of reported applications such as those to solid-state spectroscopy.
The reason for this is that light from IR/THz FEL is not so stable and the repetition rate is not so high (several Hz) despite the very high peak intensity.  
Therefore FEL is suitable as an excitation source, but the instable light intensity with a low repetition rate is not suitable for spectroscopy.  
A new-type of THz-FEL is the FELBE radiation source in Dresden-Rossendorf, Germany, which is a superconducting linear accelerator operating with a much higher pulse repetition rate of 13~MHz.~\cite{FELBE} 
Using the THz-FEL from FELBE, there have been several actual applications, such as high-field ESR combined with a pulse magnet~\cite{Zvyagin2009}, the ultrafast spectroscopy of semiconductors, microscopy using a scattering scanning near-field optical microscope (s-SNOM),~\cite{kehr-2008} and the THz spectroscopy of protein dynamics and complex materials excited by THz-FEL (THz pump--THz probe).~\cite{Bauer2012}

On the other hand，CSR has a longer wavelength than the length of the longitudinal fine structure of an electron bunch in an accelerator.~\cite{Nakazato1989}  The longitudinal fine structure then behaves like one particle with a large charge density.  Since the length of the fine structure is similar to the wavelength of THz light, CSR appears in the THz region.  The obtained CSR has $10^4$ times the intensity of ``normal'' (incoherent) SR.  The generation of CSR from a storage ring has been investigated at several SR facilities,~\cite{Carr2001,Byrd2002,Feikes2004,Takashima2005} particularly using a laser-slicing method.~\cite{Byrd2006,Shimada2007,Shimada2009}  Quasi-monochromatic CSR was also obtained owing to the periodic modulation of electron bunches created by an amplitude-modulated laser pulse at UVSOR-II.~\cite {Bielawski2008,Evain2010}  In addition, a new type of accelerator, the energy recovery linac (ERL), can generate intense CSR in the THz region, whose average power is $10^4$ to $10^6$ times higher than that of normal SR.~\cite{Carr2002}

Since CSR is a broadband source, a wide range of applications can be expected.
For example, CSR has been used to study the Josephson plasma in  high-$T_c$ cuprates~\cite{Singley2004} and the superconducting gap in boron-doped diamond.~\cite{Ortolani2006}
It has also been used for near-field microscopies.~\cite{Schade2004,Takahashi2012}
However, its use to study materials has been rather limited so far, mainly owing to the instability of its intensity.
Some application projects of CSR are being planned at present, for instance, THz pump-photoemission probe spectroscopy (TP$^3$S) to investigate the change in electronic structure after the excitation of low-energy electronic and vibrational states and quasiparticles~\cite{Kimura2010} and THz scanning near-field spectroscopy.
In Japan, a compact ERL (cERL) will be constructed at the High Energy Accelerator Research Organization, KEK and will be operated in a few years.
cERL can generate intense THz-CSR~\cite{Harada2008}, which will enable new types of experiments. 

\section{Conclusions and Outlook}\label{sec:conc}

In this paper, we have reviewed recent advances in the IR and THz spectroscopies using SR.
With high-brilliance IR/THz-SR, spectroscopy under extreme conditions such as high pressure, high magnetic field, and high spatial resolution, can be better performed than those with the conventional, thermal IR sources.
We have reviewed recent IR/THz spectroscopy and imaging studies of SCES under these conditions.
In the future, these experimental techniques are expected to be applicable to an even wider range of materials that exhibit interesting physical properties under these extreme conditions.
We have also reviewed other unique spectroscopic studies made with IR/THz-SR with a very high spatial resolution, such as microspectroscopies on micropatterned devices of graphene and individual SWCNT.
One of expected future developments with IR/THz-SR is its application to the s-SNOM technique.  
As already mentioned above, s-SNOM experiments with powerful THz-CSR have already been carried out.~\cite{Schade2004,Takahashi2012}
s-SNOM studies with the normal (incoherent) IR-SR may also offer unique research opportunities, since IR-SR covers the entire mid-IR range.
This is in contrast to most of the current s-SNOM studies where a monochromatic or narrow-band laser sources are used.~\cite{nsom-2009}
Attempts to perform  broadband, s-SNOM studies with IR-SR have already been made,\cite{Ikemoto2012,ikemoto-2008,ikemoto-2011} where near-field signals in the mid-IR have been detected with a spatial resolution much higher than that given by the diffraction limit.
In the near future, with further improvement in the signal-to-noise ratio, this technique may prove useful for studying, for example, nanoscale samples of SCES.

\section*{Acknowledgments}

We would like to acknowledge Professor Takao Nanba as a forerunner in the application of infrared synchrotron radiation.
We would also like to thank M. Matsunami, T. Iizuka, T. Nishi, H. Miyazaki, Y. S. Kwon, H. Kitazawa, K. Kanoda, H. Kimura, Y. Ikemoto, T. Moriwaki, E. Nakamura, T. Takahashi, M. Katoh, H. Sugawara, C. Sekine, A. Ochiai, H. Imai, Z. Liu, and G. L. Carr for their fruitful collaboration.
Part of this work was performed by the Use-of-UVSOR Facility Program of the Institute for Molecular Science.
Part of this work was performed at SPring-8  under the approval by JASRI (2005B0621, 2006A1186, 2007B1314, 2009A0089 through 2011A0089).
This work was partly supported by JSPS KAKENHI Grant Number 18340110, 22340107 (SK) and MEXT KAKENHI on Innovative Area ``Heavy Fermion'' Grant Number 21102512-A01 (HO).  



\begin{thebibliography}{99}
\bibitem{footnote}
Generally, the spatial resolution as well as the spot size ($\sigma$) can be defined as $\sigma = k \times \lambda/NA$, where $\lambda$ is the wavelength, $k=0.61$ is a constant from the Schwarzschild optics, and $NA$ is the numerical aperture; the focusing angle ($\sigma'$) can be written as $\sin^{-1}(NA)$.
If $NA=0.5$, the required emittance of an IR microscope can be evaluated as $\sigma\cdot\sigma'=0.63\lambda$ [m$\cdot$rad].
For instance, the required emittance of the spatial resolution of 12~$\mu$m using a Schwarzschild optic at a wavelength of 10~$\mu$m (1000~cm$^{-1}$, 0.12~eV) is $\sigma\cdot\sigma'$~=~6.4~$\mu$m$\cdot$rad or smaller.  
The emittance of a thermal IR source in a conventional FTIR is normally $\sigma\cdot\sigma'$~=~66~$\mu$m$\cdot$rad from an emission angle of 15~deg ($=2\sigma'$) and a source size of 1~mm ($=2\sigma$).
Therefore, only 1\% intensity is used for the IR microscope.
On the other hand, IR-SR is a low-emittance light source, for instance, $\sigma_x\cdot\sigma_x'$~=~6.2~$\mu$m$\cdot$rad (horizontal direction) and $\sigma_y\cdot\sigma_y'$~=~0.35~$\mu$m$\cdot$rad (vertical direction) for BL43IR of SPring-8 and $\sigma_x\cdot\sigma_x'$~=~12~$\mu$m$\cdot$rad and $\sigma_y\cdot\sigma_y'$~=~5.2~$\mu$m$\cdot$rad for BL6B of UVSOR.
Therefore, IR-SR is suitable for IR microspectroscopy because all emitted light can be focused on a sample in an IR microscope without intensity loss.
\end{thebibliography}
\end{document}